\documentclass[line]{aastex631}

	\usepackage{graphicx}	
	\usepackage{amsmath}
	\usepackage{amssymb}	
	\usepackage{longtable}
	\usepackage{threeparttable}
	\usepackage{multirow}
	\usepackage{booktabs}
	\usepackage{array}
    \usepackage[figuresright]{rotating}
    \usepackage[T1]{fontenc}

	\shorttitle{TV Col}
	\shortauthors{Sun et al.}

	\graphicspath{{./}{figures/}}

\begin{document}
			\title{A New Window for Studying Intermediate Polars and Tilted Accretion Disk Precession}


\correspondingauthor{Sheng-Bang Qian}
\email{qsb@ynao.ac.cn}	
\author{Qi-Bin Sun}
\affiliation{Yunnan Observatories, Chinese Academy of Sciences, Kunming 650216, PR of China}
\affiliation{Department of Astronomy, School of Physics and Astronomy, Yunnan University, Kunming 650091, PR of China}
\affiliation{Key Laboratory of Astroparticle Physics of Yunnan Province, Yunnan University, Kunming 650091, PR China}
\affiliation{Center for Astronomical Mega-Science, Chinese Academy of Sciences, 20A Datun Road, Chaoyang District, Beijing, 100012, PR of China}
\affiliation{University of Chinese Academy of Sciences, No.19(A) Yuquan Road, Shijingshan District, Beijing, PR of China}	

\author{Sheng-Bang Qian}

\affiliation{Department of Astronomy, School of Physics and Astronomy, Yunnan University, Kunming 650091, PR of China}
\affiliation{Key Laboratory of Astroparticle Physics of Yunnan Province, Yunnan University, Kunming 650091, PR China}

\author{Li-Ying Zhu}
\affiliation{Yunnan Observatories, Chinese Academy of Sciences, Kunming 650216, PR of China}
\affiliation{Center for Astronomical Mega-Science, Chinese Academy of Sciences, 20A Datun Road, Chaoyang District, Beijing, 100012, PR of China}
\affiliation{University of Chinese Academy of Sciences, No.19(A) Yuquan Road, Shijingshan District, Beijing, PR of China}

\author{Wen-Ping Liao}
\affiliation{Yunnan Observatories, Chinese Academy of Sciences, Kunming 650216, PR of China}
\affiliation{Center for Astronomical Mega-Science, Chinese Academy of Sciences, 20A Datun Road, Chaoyang District, Beijing, 100012, PR of China}
\affiliation{University of Chinese Academy of Sciences, No.19(A) Yuquan Road, Shijingshan District, Beijing, PR of China}

\author{Er-Gang Zhao}
\affiliation{Yunnan Observatories, Chinese Academy of Sciences, Kunming 650216, PR of China}
\affiliation{Center for Astronomical Mega-Science, Chinese Academy of Sciences, 20A Datun Road, Chaoyang District, Beijing, 100012, PR of China}

\author{Fu-Xing Li}
\affiliation{Yunnan Observatories, Chinese Academy of Sciences, Kunming 650216, PR of China}
\affiliation{Center for Astronomical Mega-Science, Chinese Academy of Sciences, 20A Datun Road, Chaoyang District, Beijing, 100012, PR of China}

\author{Xiang-Dong Shi}
\affiliation{Yunnan Observatories, Chinese Academy of Sciences, Kunming 650216, PR of China}
\affiliation{Center for Astronomical Mega-Science, Chinese Academy of Sciences, 20A Datun Road, Chaoyang District, Beijing, 100012, PR of China}

\author{Min-Yu Li}
\affiliation{Yunnan Observatories, Chinese Academy of Sciences, Kunming 650216, PR of China}
\affiliation{Center for Astronomical Mega-Science, Chinese Academy of Sciences, 20A Datun Road, Chaoyang District, Beijing, 100012, PR of China}
\affiliation{University of Chinese Academy of Sciences, No.19(A) Yuquan Road, Shijingshan District, Beijing, PR of China}		
			

			\begin{abstract}
TV Col is a long-period eclipsing intermediate polar (IPs) prototype star for the negative superhump (NSH) system. We investigate the eclipse minima, eclipse depth, and NSH amplitude based on TESS photometry. Using the same analytical method as SDSS J081256.85+191157.8, we find periodic variations of the O-C for eclipse minima and NSH amplitudes with periods of 3.939(25)\-d and 3.907(30)\-d, respectively. The periodic variation of the NSH amplitude of TV Col confirms that periodic NSH amplitude changes in response to the tilted disk precession may be universal, which is another evidence that the origin of the NSHs is related to the tilted disk precession. We suggest that the NSH amplitude variation may be similar to the superorbital signal, coming from the periodic change in visual brightness of the energy released by streams touching the tilted disk with tilted disk precession. 
Finally, we find for the first time that the eclipse depth exhibits bi-periodic variations with periods of P$ _{1} $ = 3.905(11)\-d and P$ _{2} $ = 1.953(4)\-d, respectively. P$ _{2} $ is about  half of P$ _{1} $ and the disk precession period ($\rm P_{1} \approx P _{prec}  \approx 2 \times P _{2} $). We suggest that P$_{1}$ may come from the periodic change in the brightness of the eclipse center due to tilted disk precession, while P$_{2}$ may come from two accretion curtains precessing together with the tilted disk, but more verification and discussion are necessary. The discovery of bi-periodic variations in eclipse depth provides a new window for studying IPs and tilted disk precession.
\end{abstract}



\section{Introduction} \label{sec:intro}

Cataclysmic variable stars (CVs) are semi-detached close binaries, with the primary star being a white dwarf and the secondary star being a late low-mass stellar \citep{warner1995cat}, which fills the Roche lobe and transfers material to the primary star through the Lagrangian point (L1). CVs can be categorized into different subtypes based on different characteristics. Based on their magnetic field properties, CVs can be generally categorized into two groups: magnetic CVs (MCVs), where the white dwarf has a strong magnetic field ($ > $1 MG), and weakly magnetic or non-magnetic CVs. MCVs can be subdivided into two types. If the magnetic field strength of the white dwarf is strong enough ($ > $10 MG) to synchronize the rotation period of the white dwarf with the orbital period, then the system is called an AM Herculis-type binary or Polar (due to strong optical linear and circular polarization). As the strong magnetic field of Polar disrupts the formation of the accretion disk, material from the secondary star streams along the magnetic lines of force to form accretion columns at the magnetic poles of the white dwarf \citep{1990SSRv...54..195C}. The other type of MCVs is QD Herculis-type stars or Intermediate Polars (IPs) with magnetic field strengths between those of non-magnetic CVs and polar stars,  where the magnetic field of the white dwarf is not sufficiently strong (1-10 MG) to synchronize the white dwarf rotation with the orbital period, and the accretion disk is preserved. However, the inner disk is disrupted to form two accretion curtains (\citealp{1991MNRAS.248..233H,1995ASPC...85..185H,1999ApJ...519..324H,1996PASA...13...87F}).

Dwarf novae (DNe) and novae-like stars (NLs) are subtypes of CVs belonging to non-magnetic CVs, and both have accretion disks. DNe have outbursts with amplitudes between 2-8 magnitudes and durations ranging from a few days to a few weeks, which can be explained by the thermal–viscous instability model (DIM; \citealp{lasota2001disc}) of the accretion disk. DIM holds that viscosity causes angular momentum in the disk to propagate outward and material to be transported inward through the accretion disk, but the origin of the viscosity is unspecified, and a dimensionless parameter, $ \alpha $, is usually assumed ($ \alpha $-prescription; \citealp{Shakura1973}). 
In DIM, the disk exhibits bi-stable states; the opacity changes with temperature, leading the disk to cycle through two states. When the temperature is low, the hydrogen is neutral, and the accretion disk is in a cold, stable state with low viscosity. When the temperature enables the hydrogen to be partially ionized, the opacity becomes sensitive to the temperature, and the accretion disk becomes thermally viscous and unstable. When the temperature is high enough for the hydrogen to be fully ionized, the accretion disk is in a thermally stable state with high viscosity, which is characterized by the S-type thermal equilibrium curve (See, e.g., \citealp{lasota2001disc}; \citealp{dubus2018testing}; \citealp{2020AdSpR..66.1004H} for details). NLs without outbursts, with high mass transfer rates and accretion disks in a long-term thermally stable state, are similar to the DN subtype Z Cam during brightness "standstill" \citep{Meyer1983A&A,Smak1983ApJ,Honeycutt1998PASP}.

Similarities, as well as differences, characterize the light curves of IPs and DNe; IPs also have outbursts (e.g., TV Col, \citealp{2005ASPC..330..405H}; EX Hya, \citealp{2000MNRAS.313..703H}; DW Cnc, \citealp{2008JAVSO..36...60C}), but some of them have short outburst durations and long recurrence times that cannot be explained by considering DIM alone. TV Col outbursts have been observed several times (e.g., \citealp{Szkody1984ApJ}; \citealp{1987IAUC.4508....1S}; \citealp{Augusteijn1994}; \citealp{1993MNRAS.264..132H}; \citealp{2005ASPC..330..405H}; \citealp{Bruch2022}; \citealp{Scaringi2022Natur}). \cite{2017A&A...602A.102H}'s research suggests that outburst phenomena in long-period systems with short outburst durations and long recurrence times similar to TV Col could be the result of a combination of DIM and an enhanced mass transfer from the secondary or magnetic field, with the lower mass transfer and stronger magnetic field being the result of the accretion disk being in a cold stable state, and IPs with long outburst durations can be explained by DIM. However, the recent study by \cite{Scaringi2022Natur} left the DIM and found that TV Col outburst characteristics (e.g., rapid rise, slow fall, short duration, multi-peak, and energy released) are similar to type-I X-ray outbursts, suggesting that TV Col outbursts arise from localized thermonuclear runaway events in magnetically confined accretion columns.

There are two types of hump modulations in CVs, positive superhumps (PSHs) and negative superhumps (NSHs). "Positive" or "Negative" is relative to the orbital period, and when the hump period is greater than a few percent of the orbital period, it is called PSHs, and when it is less than a few percent of the orbital period, it is called NSHs. Both hump modulations are thought to originate from the interaction between the accretion disk precession and the motion of the binary, but the physical mechanisms are quite different.

The PSH investigation originated with the discovery of a hump signal longer than the orbital period during the superoutburst of the SU UMa-type DN VW Hyi (e.g., \citealp{1974A&A....36..369V}; \citealp{1974IBVS..864....1M}; \citealp{Warner1975MNRAS.170..219W}). Different models have been proposed to explain PSH, such as the starspot model \citep{1974A&A....36..369V}, the eccentric orbit model \citep{1979MNRAS.189..293P} and the intermediate polar model \citep{1985ASIC..150..367W}, etc. Eventually, only the eccentric accretion disk model survived to be generally accepted \citep{1988MNRAS.232...35W}. Current research has shown that when the radius of the accretion disk exceeds the critical radius for the 3: 1 resonance, tidal instability is triggered, resulting in an eccentric accretion disk with a slow forward precession (e.g., \citealp{1988MNRAS.232...35W}; \citealp{lubow1991model}; \citealp{1996PASP..108...39O}). As a result of binary motion, the outer part of the eccentric disk undergoes periodic tidal stress from the secondary star, which then produces a periodic PSH (e.g., \citealp{vogt1982z}; \citealp{osaki1985irradiation}; \citealp{wood2011v344}).

The study of NSHs originated when three periodic signals (5.2 hr, 5.5 hr, and 4.0 d) were discovered by \cite{Motch1981} in TV Col, slightly shorter than the orbital period known as NSHs. Recently, more and more NSHs have been discovered with the release of large amounts of survey data (e.g., \citealp{2023MNRAS.520.3355S};  \citealp{sun2022study,2023arXiv230911033S}; \citealp{Bruch2022,Bruch2023,BruchIII}). The current study suggests that the NSHs arise from the interaction between the reverse precession of the nodal line from the tilted disk and the orbital motion of the binary (e.g., \citealp{1985A&A...143..313B}; \citealp{patterson1999permanent}; \citealp{harvey1995superhumps}). Unlike PSHs, where the accretion disk is tilted, and the precession direction is reversed in the origin of NSHs, NSHs result from the interaction of the stream with the accretion disk. When the accretion disk is tilted, the gas streams from the secondary star can enter the inner disk, resulting in the release of more energy, and the accretion disk precession with the binary star motion causes the periodic appearance of the NSHs (\citealp{Barrett1988MNRAS}; \citealp{Patterson1997PASP}; \citealp{Wood2007ApJ...661.1042W}; \citealp{Montgomery2009MNRAS}). The superorbital signal (SOR) are often found in NSH systems and are considered to be evidence of tilted disk precession, with a relationship between orbital period and NSH of $\rm 1/P_{prec}  = 1/P_{sor} = 1/P_{nsh} - 1/P_{orb}$ (e.g., \citealp{Katz1973NPhS..246...87K}; \citealp{Barrett1988MNRAS}; \citealp{harvey1995superhumps}; \citealp{wood2009sph}).

Recently, we have carried out studies of the NSHs based on survey data released by space telescopes  (\citealp{Sun2023arXiv}; \citealp{2023arXiv230905891S}; \citealp{2023arXiv230911033S}). We found that in NL SDSS J081256.85+191157.8, eclipse depth, NSH amplitude, and O - C at the eclipse minima all varied periodically in response to changes in accretion disk precession, with the periodic changes in NSH amplitude being direct evidence for the origin of NSHs from tilted accretion disk precession \citep{Sun2023arXiv}. In addition, we found that NSH amplitudes change as the outburst \citep{2023arXiv230905891S}, that NSH amplitudes decrease as the outburst rises and increase as the outburst declines, and that there is a rebound in the plateau. We subsequently verified in three DNe ASAS J1420, TZ Per, and V392 Hya outbursts that the phenomenon of NSHs changing with outbursts may not be an exception but a universal phenomenon \citep{2023arXiv230911033S}, providing a new window to study DN outbursts and the origin of NSHs.

TV Col is a classic CV with a long history of research. It is a long-period eclipsing IP. TV Col is the first CV to be identified as the optical counterpart of a hard X-ray source (2H 0526-328; \citealp{Charles1979ApJ}) and its X-rays have been observed many times in subsequent studies (e.g., \citealp{Schrijver1987}; \citealp{Ishida1995}; \citealp{Vrtilek1996}; \citealp{Ezuka1999}; \citealp{Rana2004}). The 1911 $ \pm $ 5 s period detected in X-rays by \cite{Schrijver1987} using the EXOSAT observatory was determined to be the spin period of the white dwarf. In recent studies, \cite{Rana2004} has more accurately determined the X-ray period to be 1909.67 $ \pm $ 2.5 s based on the Rossi X-Ray Timing Explorer (RXTE) Proportional Counter Array. In addition, \cite{1985A&A...143..313B} claimed the observation of a 1938 $ \pm $ 10 s periodic signal in the optical bands (UBV). \cite{Bruch2022} discovered a beat (40.878 (2) d$ ^{-1} $) of white dwarf spin signal to the orbital signal based on TESS data, also discovered by \cite{Scaringi2022Natur}, confirming the existence of an optical band of the spin signal.

TV Col was the first object found to have a NSH signal and can be called the prototype star of the NSH system. The research on the NSH originated from \cite{Motch1981}'s discovery of three periodic signals (5.2 hr, 5.5 hr, and 4.0 d) in the TV Col. The three cycles are certified from shortest to longest as the NSH period, the orbital period and the accretion disk precession period. Numerous studies have been accurate on these three signals (e.g., \citealp{Hutchings1981ApJ}; \citealp{Barrett1988MNRAS}; \citealp{Hellier1991MNRAS}; \citealp{Augusteijn1994}; \citealp{Bruch2022}). In addition \cite{2003MNRAS.340..679R} found the presence of PSH with 0.2639(35) d in TV Col. Based on a large amount of TESS data, \cite{Bruch2022} determined the NSH, orbital period and precession period to be 0.22860010(2) d, 0.215995(1) d and 3.895(5) d, respectively, but no PSHs were found.

In conclusion, TV Col is a very special CV, and the study of TV Col contributes to the knowledge and understanding of various phenomena of CVs, especially the unique components of IPs and tilted disk precession and the origin of NSHs. In this paper, based on the TESS photometry, we have investigated the NSH, SOR, eclipse depth, and eclipse minima of TV Col. The purpose is to verify the effects brought by the accretion disk precession on various aspects of the CVs' characteristics and to try to discover new observational phenomena. The paper is organized as follows.
Section \ref{sec:style} presents the \textit{TESS} photometry.
Section \ref{sec:Analysis} is the NSH amplitude, eclipse depth, and O-C analysis.
Possible reasons for variations in eclipse depth and NSH amplitude are discussed in Section \ref{sec:DISCUSSION}.
Section \ref{sec:CONCLUSIONS} is the summary.

\begin{figure}
\centerline{\includegraphics[width=0.7\columnwidth]{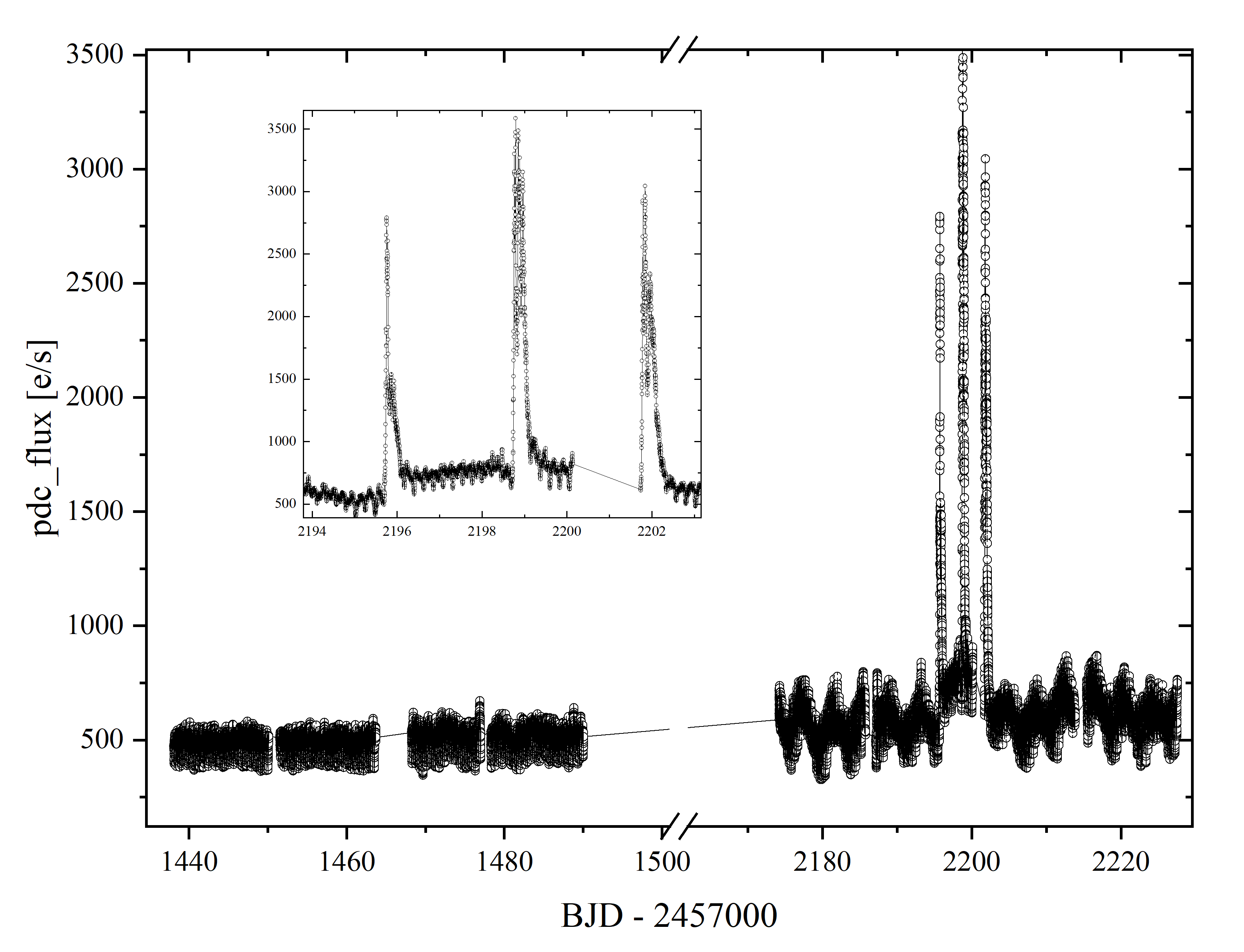}}
\caption{TV Col light curve from TESS photometry. The subfigure is an enlarged view of the outburst light curve.\label{light}}
\end{figure}


\section{DATA} \label{sec:style}

In this paper, we use the space telescope \textit{Transiting Exoplanet Survey Satellite} (\textit{TESS}, \citealp{Ricker2015journal}) survey data, downloaded from the Mikulski Archive for Space Telescopes (MAST)\footnote{https://mast.stsci.edu/}. TESS is a space-based optical all-sky survey telescope launched on April 18, 2018, as a Massachusetts Institute of Technology (MIT)-led NASA mission to search for exoplanets. However, the TESS survey data release contains many photometric data from CVs, providing a valuable opportunity to study CVs. TESS divides the sky into different sectors, each of which is observed for about 27.4 days in the wavelength range of 600 to 1000 nm, with an exposure time of 2 minutes in the short cadence mode, which provides data support for the study of the evolution of light curves on long timescales of CVs. We have already carried out partial work on CVs based on TESS data (e.g., \citealp{2023arXiv230905891S}; \citealp{Sun2023arXiv}; \citealp{Sun2023MNRAS}; \citealp{2023arXiv230911033S}; \citealp{sun2022study}).

TV Col was photometered by TESS in sectors 5 (MBJD 58437 (2018-11-15) - MBJD 58463 (2018-12-11)), 6 (MBJD 58467 (2018-12-15) - MBJD 58489 (2019-01-06)), 32 (MBJD 59173 (2020-11-20) - MBJD 59199 (2020-12-16)) and 33 (MBJD 59201 ((2020-12-18) - MBJD 59227 (2021-01-13)), respectively (see Figure. \ref{light}).
\cite{Scaringi2022Natur} and \cite{Bruch2022} have used the same TESS data to discover the NSHs, the orbital period, and the accretion disk precession period, as well as the harmonics of the white dwarf rotation period with the orbital period, also analyzing the outbursts, but has not delved into the details of the variations in the TV Col. \cite{Scaringi2022Natur} focused mainly on TV Col outburst while \cite{Bruch2022} focused on frequency analysis. This paper differs significantly from previous studies in that we will focus on the effects of accretion disk precession on TV Col in various aspects, such as changes in NSH amplitude, changes in eclipse minima, and changes in eclipse depth.

\begin{figure}
	\centerline{\includegraphics[width=0.7\columnwidth]{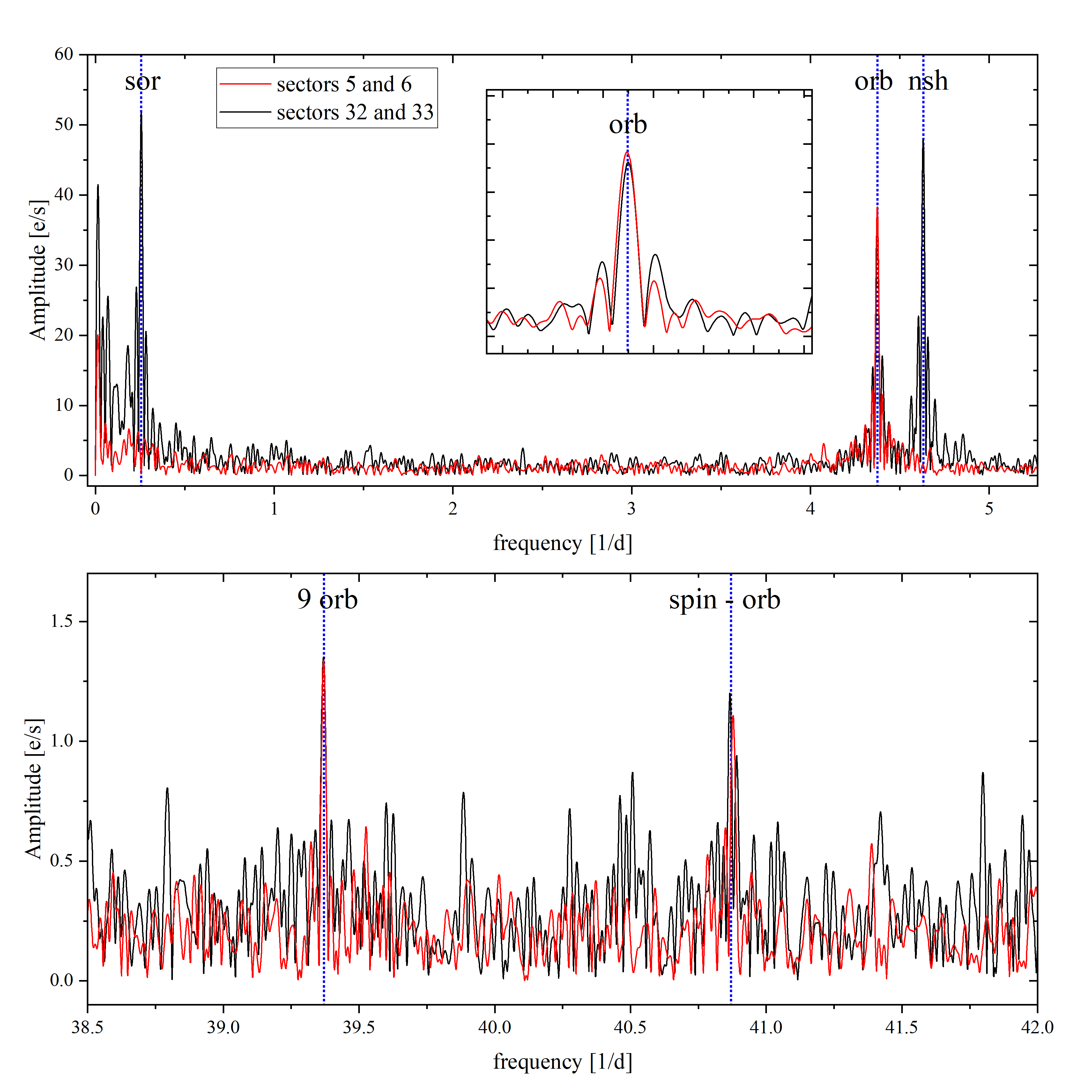}}
	\caption{Frequency - Amplitude spectrum of TV Col. The red curves are the results for sectors 5 and 6, and the black curves are for sectors 32 and 33.\label{fft}}
\end{figure}


\section{Analysis and Results} \label{sec:Analysis}

\cite{Scaringi2022Natur} and \cite{Bruch2022} have performed two frequency analysis. They found orbital signals and harmonics of the spin and orbital signals in sectors 5, 6, 32, and 33, but NSH and SOR were only present in sectors 32 and 33. Because we need to discuss the origin of the NSH, we will perform the frequency analysis in our work. We split the data into two parts, but the difference is that we removed the three outbursts in sectors 32 and 33. The two sections of data were frequency analyzed separately using \texttt{Period04} software (see \citealp{Period04} for details), and for comparison, we plotted the two spectrograms together, and the results are essentially the same as \cite{Scaringi2022Natur} and \cite{Bruch2022}, it can be found that the harmonics of the spin signal and the orbital signal are present in both parts of the data (see Figure. \ref{fft}).

\subsection{Calculate the NSH amplitude} \label{subsec:amplitude}

\begin{figure}
\centerline{\includegraphics[width=0.85\columnwidth]{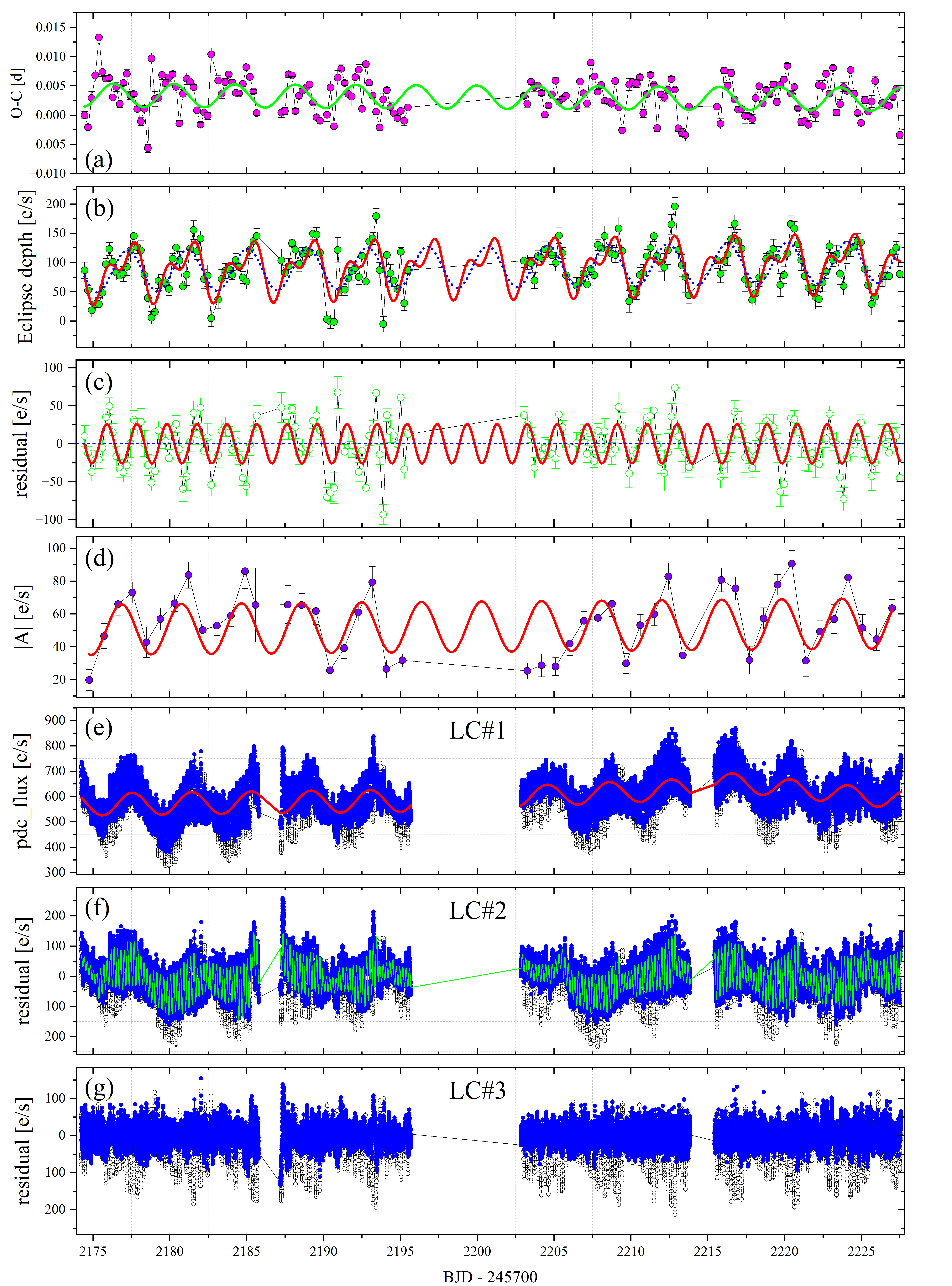}}
\caption{Different curves and analytical procedures for TV Col. (a): O-C plots, the solid green line is a linear superimposed sinusoidal fit; (b): Eclipse depth curves, blue dashed line and red solid line are single and double sine fit fits, respectively; (c) Residuals from the single sine fit in plate b, and the solid red line is the single sine fit; \textbf{(d)}: plots of the absolute value of the NSH amplitude; (e): LC\#1 light curve, where the outburst was removed from the original curve, the solid blue circle is the out-of-eclipse part, the red curve is a linear superimposed sine fit to the out-of-eclipse curve; (f): LC\#2 light curve, residuals from the fit to LC\#1 in plate e, the green curve is a segmented linear superimposed sine fit to the out-of-eclipse curve; (g) LC\#3 light curve, residuals of the segmented linear superposition sinusoidal fit of LC\#2 in plate f.\label{4}}
\end{figure}

In our recent work \citep{Sun2023arXiv}, we have found that the eclipse minima, the eclipse depth and the NSH amplitude vary periodically with the accretion disk precession in SDSS J081256.85+191157.8, and in order to verify the generality of these phenomena we plan to explore if there are similar variations in TV Col, a prototype star of the NSH system. Therefore, we adopted a processing method for the light curve of TV Col that is essentially the same as SDSS J081256.85+191157.8.
The NSHs and accretion disk precession signals for TV Col are present only in sectors 32 and 33, so we first analyzed the variation of the NSH amplitude in sectors 32 and 33, using the data after removing the outburst (see Figure. \ref{4}e; we named the light curve as LC\#1). We will perform the following steps:

(i) We split the LC\#1 into three parts, and then the least-squares linear superimposed sine model without weights is used to fit the out-of-eclipse curves (the solid blue circle for curve LC\#1 in Figure. \ref{4}e):
\begin{equation} \label{eq:sine}
	\rm Flux(BJD) = Z + B*BJD + A*\sin(2\pi*(freq*BJD+p))
\end{equation}
$\rm Z$, $\rm B$, $\rm A$, $\rm freq$, and $\rm p$ are the fitted intercept, slope, amplitude, frequency, and phase, respectively. The fitted curve was then subtracted from all curves to obtain the final light curve with the precession signal removed. The light curve with the precession signal removed is named LC\#2 (see Figure. \ref{4}f).

(ii) The out-of-eclipse curve of LC\#2 was divided into 47 segments of about 0.9139 days each (4 $ \times $ P$ _{\rm orb} $; some segments are in the vicinity of large data gaps of slightly longer or shorter duration), and a linear superimposed sine was fitted to each segment. The final light curve with all curves of LC\#2 minus the fitted curve to obtain the light curve with NSH removed was named LC\#3 (see Figure. \ref{4}g).

Consistent with the analytical method of SDSS J081256.85+191157.8, we use segments for linear superposition sinusoidal fitting to obtain amplitude information for each NSH segment, and the best fits are shown in Table \ref{tab:A1}. The average BJD for each segment was plotted as the x-value, and the absolute amplitude value obtained from the fit was plotted as the y-value. The results are shown in Figure. \ref{4}d. Although there is a poor number of data points, it is still possible to find periodic variations in the NSH amplitudes, which were fitted using Eq. \ref{eq:sine} to obtain the best fitting frequency of 0.256(2) d$ ^{-1} $ (3.907(30) d) (see Table. \ref{tab:1}). It can be found that TV Col is consistent with SDSS J081256.85+191157.8 in that there is a periodic variation in the NSH amplitude similar to the accretion disk precession cycle.

\begin{figure}
\centerline{\includegraphics[width=0.8\columnwidth]{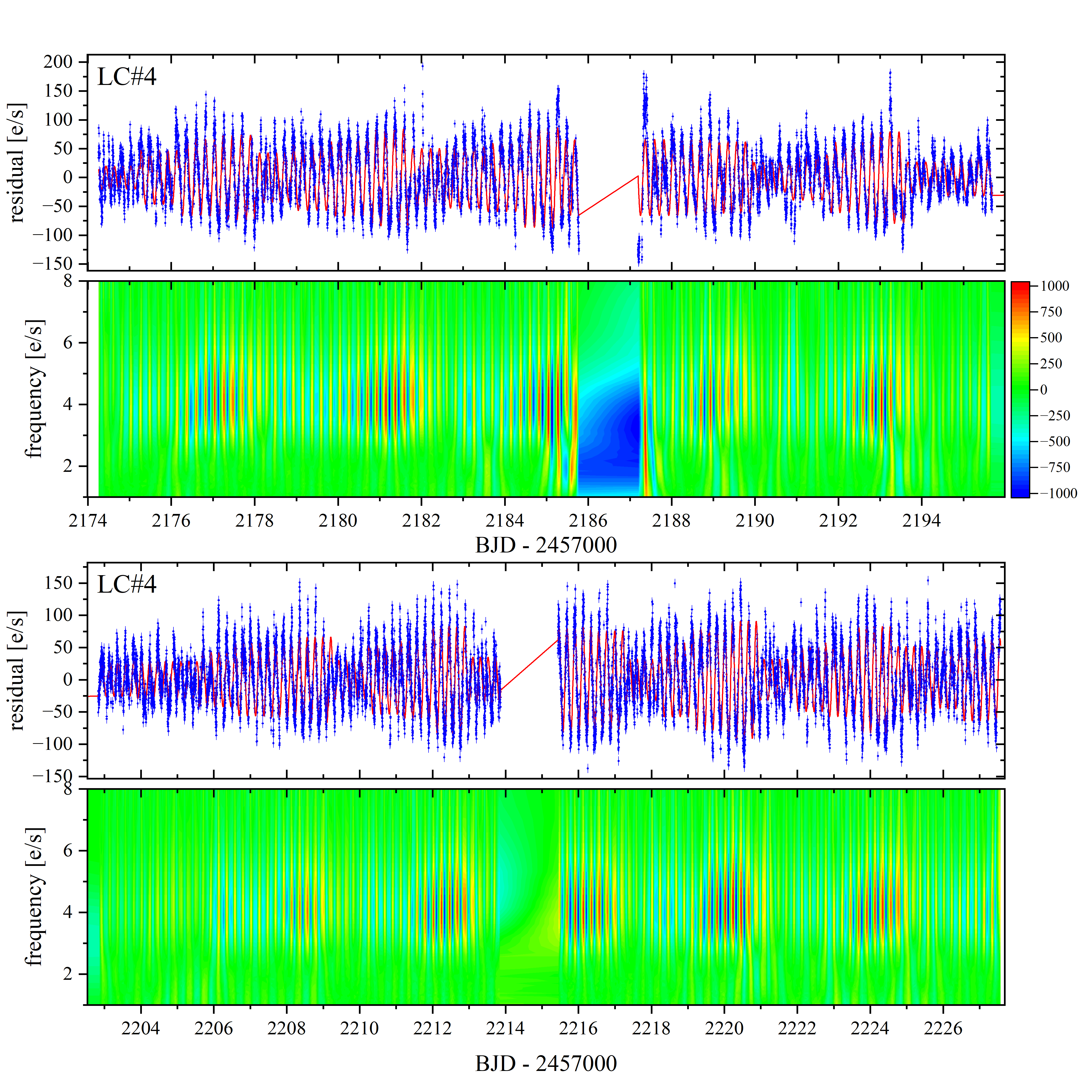}}
\caption{LC\#4 light curve and the corresponding 2D CWT spectrum. LC\#4 is an out-of-eclipse curve of LC\#2 with the linear part of the segmented linear superposition sinusoidal fit removed. The red solid line is the green solid line in Figure. \ref{4}f with the linear part removed.\label{CWT}}
\end{figure}

We will use the Continuous Wavelet Transform (CWT; \citealp{CWT1989}) to demonstrate more visually the periodic variation of the NSH amplitude. For the out-of-eclipse curve of LC\#2, we remove only the linear part obtained from the segmented linear superposition sinusoidal fit (see the blue dotted line in Figure. \ref{CWT}, named LC\#4). Similarly, only the sinusoidal part is retained in the theoretical curve obtained from the fit (see the red solid line in Figure. \ref{CWT}), where the naked eye can find a periodic variation of the NSH amplitude. The two-dimensional (2D) power spectrum obtained using CWT for LC\#4 reveals the same periodic variation in the 2D CWT power spectrum, with the stripes changing periodically as the NSH amplitude changes (see Figure. \ref{CWT}).

\subsection{O-C analysis} \label{subsec:O-C}

Using a method consistent with \citep{Sun2023arXiv}, we used a Gaussian fit to compute the eclipse minima for LC\#3, which ultimately obtained 182 minima. We use the first minima as the initial epoch, 0.22860010(2) d as the period obtained by \cite{Bruch2022}, to obtain the ephemeris:

\begin{equation} \label{eq:ephemeris}
\rm	Min.I = BJD2459174.4534(6) + 0.22860010(2) \times E 
\end{equation}
E is the number of circles. The O-C analysis of all the minima used this ephemeris, and the results suggest that there may be a periodic variation in O-C (see Figure. \ref{4}a and Table \ref{tab:A2}). We obtained spectrograms using Period04 and determined that the frequency and amplitude of the periodic variations were 0.2538(16) d$ ^{-1} $ (3.939(25) d) and 0.0018(3) d ($\sim$ 155 s) (see Figure. \ref{2FFT}b), respectively, using Eq. \ref{eq:sine} to fit the O-C, and the results were consistent with the frequency analysis (see Figure. \ref{4}a and Table \ref{tab:1}). Consistent with SDSS J081256.85+191157.8, periodic variations in the O-C of TV Col are similar to the accretion disk precession cycle. Both \cite{Miguel2016accretion} (UX UMa) and \cite{Sun2023arXiv} (SDSS J081256.85+191157.8) suggested that this change could be due to a change in the center of brightness as a result of the tilted accretion disk precession.

\subsection{Analysis of eclipse depth} \label{subsec:depth}

The light curve of LC\#3 has been fitted by two linear superimposed sinusoidal fits on the original light curve as far as possible to exclude the interference of the NSH and SOR signals in calculating the out-of-eclipse curve. Therefore, similar to SDSS J081256.85+191157.8, we used the average fluxes of the -0.5 to -0.1 phases and 0.1 to 0.5 phases as the out-of-eclipse light curves corresponding to each eclipse (see \cite{Sun2023arXiv}'s Figure. 4), and finally calculated the eclipse depth based on the minima and the average of the out-of-eclipse curves. The final eclipse depth variation curve obtained is shown in Figure. \ref{4}b, where an apparent periodic variation can be found. Based on the frequency analysis, we obtain two periodic signals f$ _{1} $ = 0.2561(7) d$ ^{-1} $ (3.905(11) d) and f$ _{2} $ = 0.5122(10) d$ ^{-1} $ (1.953(4) d), both of which are close to a double relationship. 

Fitting the eclipse depth curve using Eq. \ref{eq:sine} revealed that a single sinusoidal change could not describe the variation (see the blue dotted line in Figure. \ref{4}b). Therefore, we used a linear superposition bi-sinusoidal model to verify the variation of eclipse depth:
\begin{equation} \label{eq:sine2}
\rm Flux(BJD) = Z + B*BJD + A_{1}*\sin(2\pi*(freq_{1}*BJD+p_{1})) + A_{2}*\sin(2\pi*(freq_{2}*BJD+p_{2}))
\end{equation}
This equation adds an extra sinusoidal term to Eq. \ref{eq:sine}. The bi-sinusoidal fit ($\rm \chi_{red}^2$ = 3.65) is better than the single-sine fit ($\rm \chi_{red}^2$ = 5.49), indicating that f$ _{2} $ is not a harmonic of f$ _{1} $. The best fits are shown in Table \ref{tab:1}, and the best-fit frequencies of 0.2563(8) d$ ^{-1} $ and 0.5125(10) d$ ^{-1} $ are consistent with the results of the frequency analysis, further revealing the existence of a bi-periodic variation in the eclipse depth, with a doubling relationship in the period of change. 
Although the O-C curve also had a signal around 3.895(5)/2 d, the introduction of a bi-sinusoidal fit did not significantly optimize the fit, and the dispersion of the data was such that we did not consider the bi-sinusoidal variation further.

\subsection{Light curves for sectors 5 and 6} \label{subsec:56}

\begin{figure}
\centerline{\includegraphics[width=0.6\columnwidth]{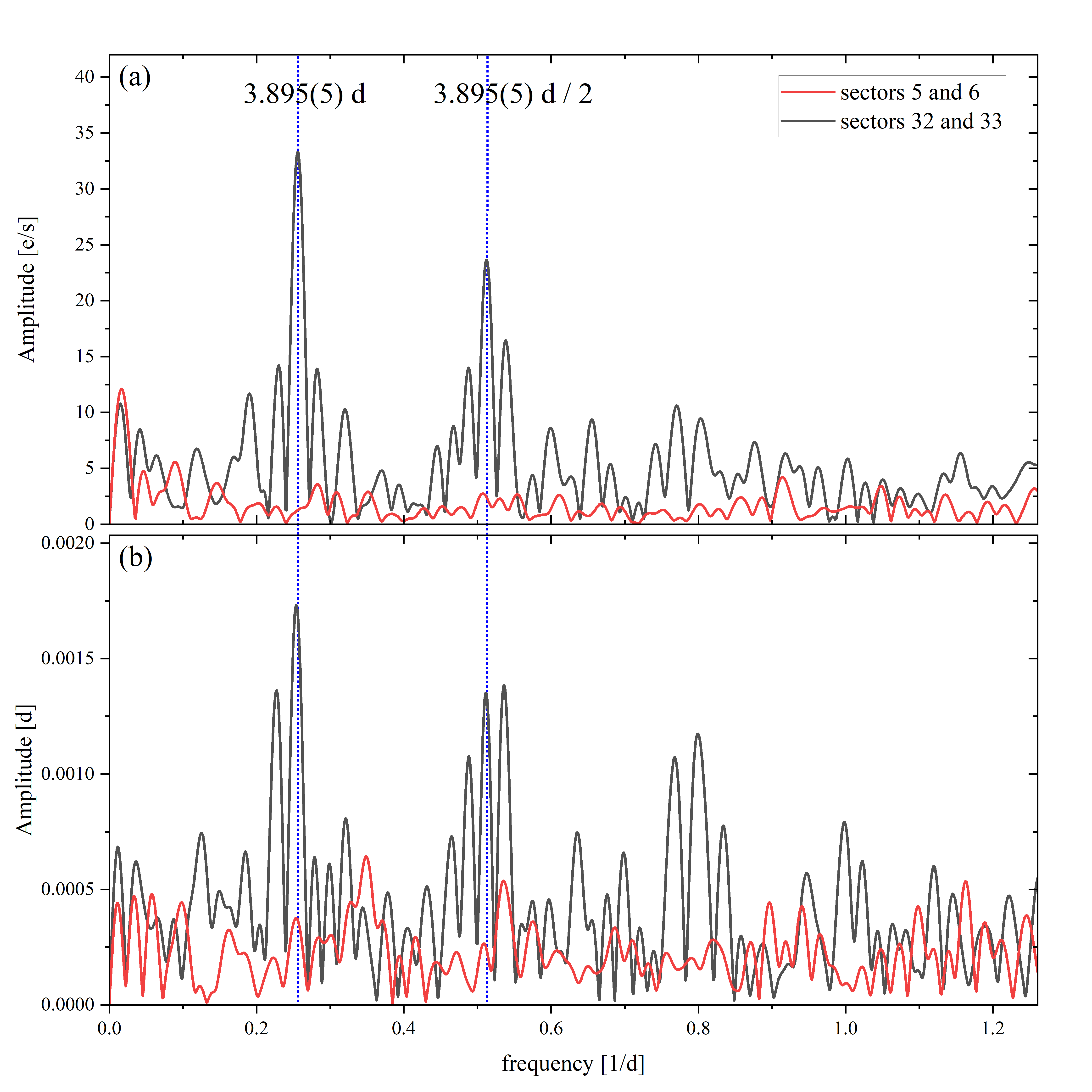}}
\caption{Frequency-Amplitude spectrum of eclipse depth and O-C curves. (a) spectrum of eclipse depth, the red and black solid lines represent data from different sectors; (b) spectrum of O-C curve. The two blue vertical lines correspond to periods of 3.895(5) d and 3.895(5)/2 d.\label{2FFT}}
\end{figure}

\begin{figure}
\centerline{\includegraphics[width=0.6\columnwidth]{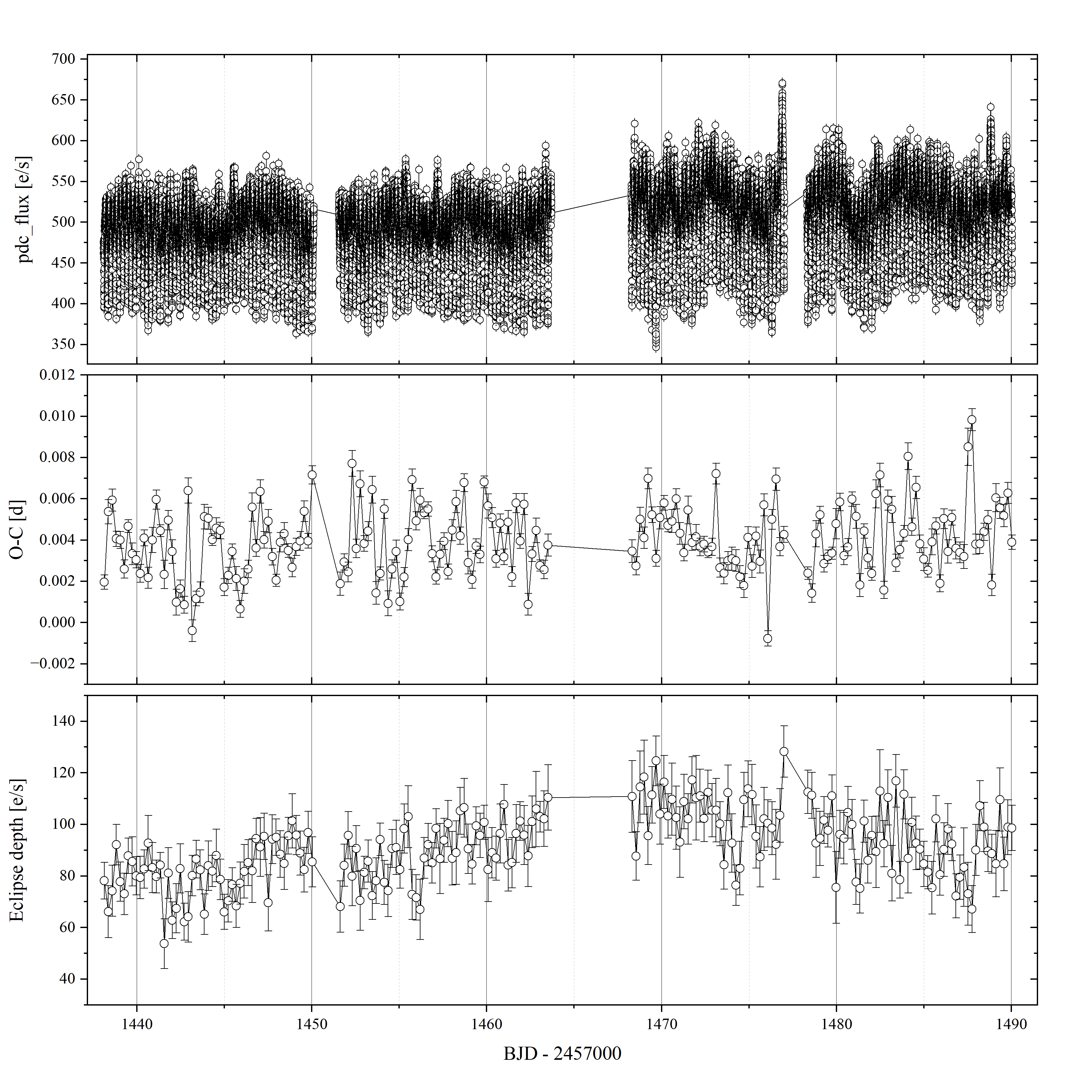}}
\caption{Light curves, O-C plots, and eclipse depth plots for sectors 5 and 6 of TV Col.\label{sector5_6}}
\end{figure}

No NSH and SOR signals were found in sectors 5 and 6, and for comparison with the calculations for sectors 32 and 33, we calculated the changes in O-C and eclipse depth using the same methods (see Figure. \ref{sector5_6}). The frequency analysis results indicate that the O-C and eclipse depth in sectors 5 and 6 do not show similar cyclic variations as in sectors 32 and 33 (see Figure. \ref{2FFT}).

\begin{figure}
\centerline{\includegraphics[width=0.8\columnwidth]{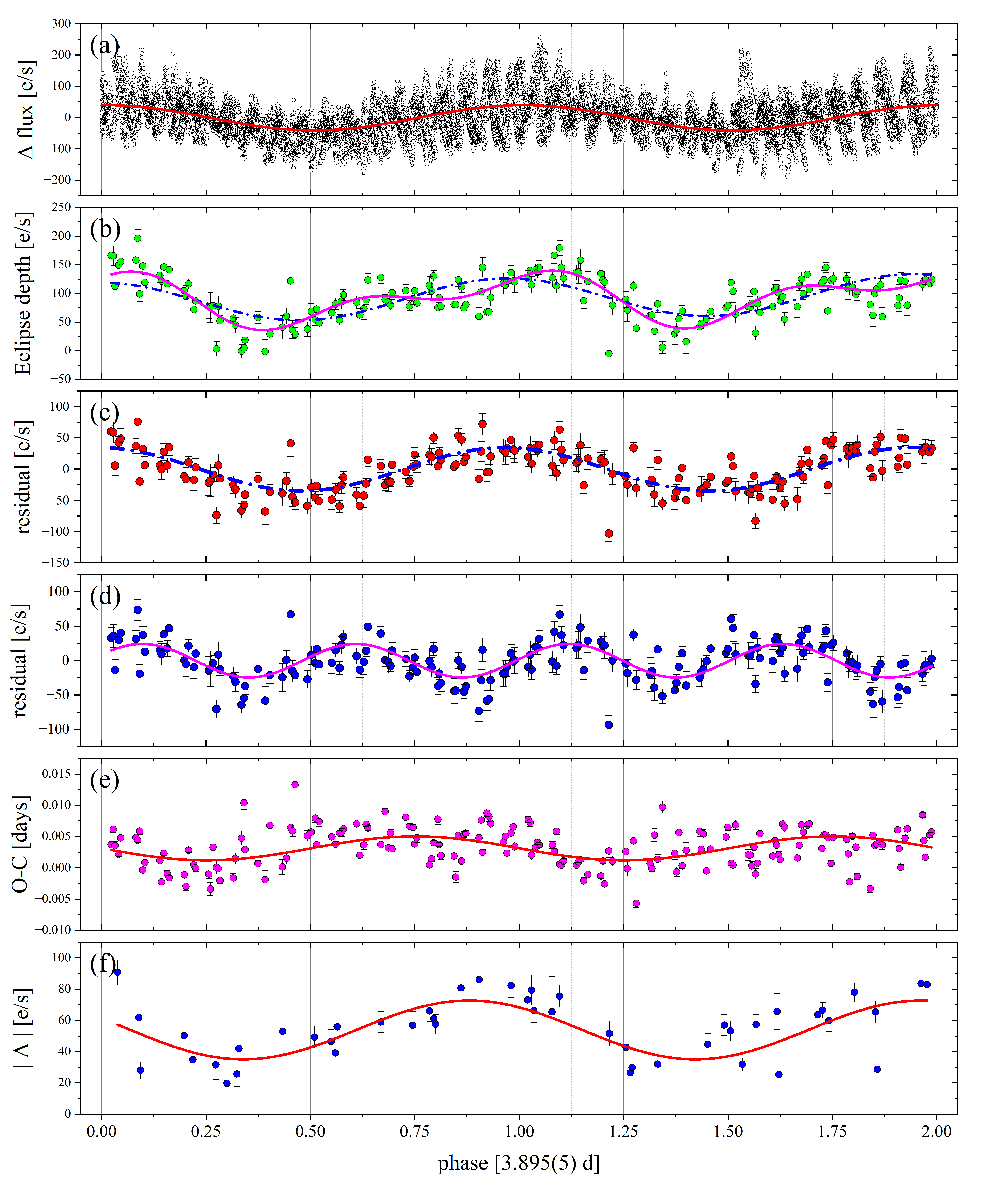}}
\caption{Different folding curves for TV Col. (a): the out-of-eclipse curve's folded curve after removing the linear part of LC\#1 in Figure. \ref{4}e, the red curve is a sinusoidal fit; (b): Folded curves of eclipse depth, blue and magenta lines are single and double sine model fits; (c) folded curve of the eclipse depth curve after removing the linear and sinusoidal terms with period P$ _{2} $; (d), (e) and (f) are folded curves of the eclipse depth, O-C and the absolute values of NSH amplitudes, respectively, corresponding to Figure. \ref{4}c, a and d.\label{flod}}
\end{figure}

\begin{deluxetable*}{cccccccccccccccc}
\tablenum{1}
\tablecaption{Sinusoidal fitting results for different curves.\label{tab:1}}
\setlength{\tabcolsep}{0.01mm}
\tiny
\tablehead{
	\colhead{Name} &	\colhead{Start$^{\rm a}$} & \colhead{End$^{\rm a}$} &  \colhead{During} & 
	\colhead{Z} & \colhead{Error} & 
	\colhead{B} & \colhead{Error} & 
	\colhead{A} & \colhead{Error} &
	\colhead{freq} & \colhead{Error} &
	\colhead{p} & \colhead{Error} &
	\colhead{$\rm \nu^{ b}$} &\colhead{$\rm \chi_{red}^2$$^{c}$}}
\startdata
SOR&2174.25811 &2195.69022 &21.43 &-954.138 &185.144 &0.701 &0.085 &-43.948 &0.759 &0.25796 &0.00043 &-34.966 &0.943 &10967&3183.638 \\
&2202.83184 &2213.86639 &11.03 &-4979.470 &472.590 &2.533 &0.214 &41.701 &0.970 &0.24749 &0.00108 &59.654 &2.394 &6048&2670.681 \\
&2215.43720 &2227.57437 &12.14 &14419.020 &458.574 &-6.209 &0.206 &-36.377 &0.994 &0.26682 &0.00119 &-232.683 &2.644 &6637&3059.721 \\
O - C&2174.45338 &2227.48525 &53.03 &0.033 &0.026 &-1.34E-05&1.18E-05&1.99E-03&1.54E-03&0.25365 &0.00127 &0.220 &0.153 &177&23.668 \\
A&2174.74476 &2226.99608 &52.25 &-107.791 &303.631 &0.073 &0.138 &15.319 &14.815 &0.25598 &0.00196 &1.020 &0.223 &42&4.971 \\
Eclipse depth $^{\rm d}$&2174.45338 &2227.48525 &53.03 &-603.038 &244.678 &0.316 &0.111 &35.593 &12.778 &0.25631 &0.00078 &0.681 &0.083 &174&3.650 \\
&&&&&&&&25.863 &16.841 &0.51254 &0.00099 &0.842 &0.047 &&\\
\enddata
\begin{threeparttable} 
	\begin{tablenotes} 
		\item [a] BJD - 2457000. 
		\item [b] Degree of freedom of the fit. 
		\item [c] Fitted Reduced Chi-Square.
		\item [d] The eclipse depth was fitted using Eq. \ref{eq:sine2} with two sinusoidal terms.
	\end{tablenotes} 
\end{threeparttable}		
\end{deluxetable*}

\section{DISCUSSION} \label{sec:DISCUSSION}

\subsection{Accretion curtains?} \label{subsec:NSH and SOR}

In NL AQ Men with NSHs, \cite{2013MNRAS.435..707A} first focused on variations in eclipse depth, suggesting that eclipse depth could be a window to study tilted disk precession, but no quantitative analysis was performed. Subsequently, \cite{Boyd2017} and \cite{2021MNRAS.503.4050I} found periodic variations in eclipse depth close to the accretion disk precession cycle in NL DW UMa and NL AQ Men, respectively. In recent work \citep{Sun2023arXiv}, we have similarly found periodic variations in the eclipse depth of SDSS J081256.85+191157.8 close to the accretion disk precession cycle. These works confirm that variations in eclipse depth with tilted disk precession are common, but we found a bi-periodic variation in TV Col that differs from previous studies and may have significant implications for investigating tilted disk precession and CVs.

In the current work, we found a new phenomenon of eclipse depth variation, with a bi-periodic variation in the eclipse depth of TV Col with periods of P$ _{1} $ = 3.905(11) d and P$ _{2} $ = 1.953(4) d. 
P$ _{1} $ is twice as large as P$ _{2} $ approximately equal to the period of the tilted disk precession ($ \rm P_{1} \approx P _{prec}  \approx 2 \times P _{2} $). P$ _{1} $ can be interpreted as a periodic variation brought about by tilted disk precession, but the appearance of P$ _{2} $ proves that there are other eclipse centers of the accretion disk, and they appear twice in one cycle of the tilted disk precession.
As in SDSS J081256.85+191157.8, we used a maximum of the SOR as the zero-phase point ($\rm \phi_{prec}$ = 0), and used 3.895(5) d as the period to fold the out-of-eclipse light curve of LC\#1 with the linear part removed (see Figure. \ref{flod}a). The O-C curves, the eclipse depth curves, and the NSH amplitudes were also folded using the same method (see Figure. \ref{flod}a, b, e, and f). The folded eclipse depth curves were fitted using a bi-sinusoidal model, and the best fit is shown in the magenta curve in Figure. \ref{flod}b. We split the fitted and folded curves into Figures \ref{flod}c and d to demonstrate the two periodic variations. Figures \ref{flod}c shows the folded curves in Figure. \ref{flod}b with the sinusoidal of period P$ _{2} $ and linear component removed, and the blue curve is the corresponding fitted curve; Figure. \ref{flod}d shows the sinusoidal component of P$ _{1} $ and linear removed; the magenta curve is the corresponding fitted curve. The results show that the component of the period variation of the eclipse depth with a period of P$ _{1} $ = 3.905(11) d is maximum and minimum at $\rm \phi_{prec} \approx $ 0 and $\rm \phi_{prec} \approx $ 0.5, respectively. The component with a period of  P$ _{2} $ = 1.953(4) d is maximum at $\rm \phi_{prec} \approx $ 0.125 and $\rm \phi_{prec} \approx $ 0.625. The folded curves more visually demonstrate the bi-periodic variation in eclipse depth. The periodic variation with a period of  P$ _{1} $ = 3.905(11) d reaches a maximum at $\rm \phi_{prec} \approx $ 0, which suggests that the origin of this component comes from the tilted disk precession resulting in a periodic variation of the tilt disk brightness center. 

The presence of periodic variations in the eclipse depth of TV Col with a period of P$ _{2} $ =  1.953(4) d about half the period of the tilted disk precession is the most significant difference from AQ Men, DW UMa, and SDSS J081256.85+191157.8 and is a new phenomenon in the study of the tilted disk precession and CVs. We will discuss the origin of the variation in eclipse depth with a period of P$ _{2} $ in the context of the observational characterization of TV Col:

(a) The change in eclipse depth was only present in sectors 32 and 33 and disappeared in sectors 5 and 6 (no NSH and SOR were detected), which proves that the change in eclipse depth is related to the tilted disk precession.

(b) Changes in eclipse depth characterize changes in the brightness centers of the primary star system (white dwarf, hotspots, streams, and accretion disk). Because P$ _{2} $ is about half of the period of tilted disk precession, this proves that the source of P$ _{2} $ is associated with the accretion disk.

(c) The signal of the beat of the white dwarf spin with the orbital period is found in sectors 5, 6, 32, and 33, but the variation of the eclipse depth is only present in sectors 32 and 33, which proves that it is not necessarily linked to the white dwarf spin.

(d) AQ Men, DW UMa, and SDSS J081256.85+191157.8 all belong to the NL and are non-magnetic CVs with thermally stable state accretion disks and periodic variations in eclipse depth similar to the tilted disk precession cycle, but variations of about half the precession cycle were not found. This demonstrates the uniqueness of TV Col. The difference from the three stars is that TV Col is an IP with a stronger magnetic field and a truncated inner disk but a combined structure of an accretion disk and two accretion curtains. This demonstrates that the variation in eclipse depth in TV Col with a period of P$ _{2} $ may be associated with either the magnetic field or the accretion curtains.

Therefore, the variation in eclipse depth with period P$ _{2} $ =  1.953(4) d can only be related to the accretion disk and the unique characteristic of IPs, and we suggest that this variation may be caused by the two accretion curtains precession together with the tilted disk. In IPs, the inner disk will be truncated by the magnetic field, and the falling material from the accretion disk will flow along the magnetic field lines to magnetic poles of the white dwarf, forming two arc-shaped accretion curtains (see \citealp{1991MNRAS.248..233H,1996PASA...13...87F,1999A&A...347..203N} for details). If two accretion curtains are precession together with the tilted disk, two will emerge in a single precession cycle, consistent with the periodic variation in eclipse depth half the precession cycle.

However, it is important to note that our explanation presupposes that the accretion disk does not undergo shape changes in the vertical direction and that the accretion curtains can be eclipsed. The change in eclipse depth with period P$ _{2} $ may also be caused by variations in brightness and shape of the tilted disk, which precess with the tilted disk. It is likely that these changes are related to the properties of the IPs, but the relevant theory is lacking.

An alternative perspective is that we may be over-interpreting the change at P$ _{2} $. It is also possible that no physical component of the system is changing at P$ _{2} $. Instead, the only variations are at the precessional period P$ _{1} $, but the variation at the P$ _{1} $ precessional period has a non-sinusoidal profile. That would then produce a harmonic of P$ _{1} $ (namely half-P$ _{1} $ = P$ _{2} $) in the periodogram.

Therefore, our suggestion needs to be tested and discussed in more follow-up studies. 
We strongly recommend the use of smoothed particle hydrodynamic (SPH) simulations similar to those employed in the studies of \cite{Wood2000ApJ}, \cite{wood2009sph}, and \cite{Montgomery2012}, and then considered different ingredients to test the observations we obtained.

\subsection{Variations in the NSH amplitude} 

We have again found periodic variations of the NSH amplitude with tilted disk precession in TV Col, a prototype star of the NSH system, which demonstrates that this phenomenon may be universal and provides new evidence to study the origin of NSHs. In SDSS J081256.85+191157.8, we have found that the NSH amplitude varies periodically with the tilted disk precession \citep{Sun2023arXiv}, but there is no answer as to what causes the NSH amplitude to vary.

Back to the origin of the NSH, due to the reverse precession of the tilted accretion disk, streams from the secondary star will have the opportunity to enter the inner disk and thus release as much energy as possible (\citealp{Barrett1988MNRAS}, \citealp{Patterson1997PASP}, \citealp{Wood2007ApJ...661.1042W}). First, it is assumed that the accretion disk is tilted but not precession; after the secondary star has moved in a circle around the primary star, streams will have the opportunity to enter the inner disk twice at the farthest point, producing two humps with periods half of the orbital period; because the disk is optically thick and tilted, the observer will only see one hump during a circle \citep{Montgomery2012}. If the accretion disk is tilted and reverse precession, this results in a hump period that is less than the orbital period, coinciding with the observed NSH period.

The new phenomenon is that the NSH amplitude changes periodically with the tilted disk precession, which proves that this phenomenon is related to the tilted disk precession. 
In the NSH origin theory, the impact point between streams and disk varies periodically, but the variation period is the NSH cycle instead of the tilted disk precession cycle. If the period of change of the impact point of streams and accretion disk is the precession period, the resulting hump period is consistent with the precession period, which is inconsistent with the observation. In addition, the trajectory of the impact point of streams and accretion disk is stable whether the tilted disk is precession or not, so we excluded NSH amplitude variations from variations with the position of the tilted disk in contact with streams. The NSH amplitude reaches its maximum and minimum at $\rm \phi_{prec} \approx $ 0.875 and $\rm \phi_{prec} \approx $ 0.375 (see Figure. \ref{flod}f), respectively, which is a strong indication that the NSH amplitude variation is correlated with the phase of the tilted disk precession. The tilted disk precession leads to a periodic change in the visual area of the tilted disk, which is reflected as a SOR signal on the light curve, suggesting that the periodic change in the NSH amplitude may similarly come from the periodic change in the visual brightness of the energy released by the stream impinging on the disk surface with the different phases of the tilted disk precession. This is new evidence that the NSH originated from the interaction of streams from a secondary star with the reverse precession of a tilted accretion disk.


\section{CONCLUSIONS} \label{sec:CONCLUSIONS}

This paper investigates a long-period eclipsing IP TV Col, the prototype star of NSHs, based on TESS photometry. We find that the NSH amplitude and the O - C of the eclipse minima have periodic variations similar to those of the tilted accretion disk precession. In addition, we find a bi-periodic variation in the eclipse depth of TV Col.

Based on the segmented linear superposition sinusoidal fitting, we find that there is a periodic variation of the NSH amplitude of TV Col with a period of 3.907(30) d, which is close to the period of the tilted disk precession $ (\rm P _{prec} $ = 3.895(5) d), and this result is verified by the CWT and folding curve. The SOR signal comes from the change in the visual area of the accretion disk due to the tilted disk precession. Therefore, we suggest that the periodic variation of the NSH amplitude may be similar to the SOR signal, and that the energy released by streams from secondary stars touching the tilted disk may similarly periodically vary in visual brightness with the tilted disk precession. This is new evidence that the NSH originates from the interaction between streams and the reverse precession of the tilted accretion disk.

In addition, we find that the eclipse depth of TV Col varies with periods of  P$ _{1} $ = 3.905(11) d and P$ _{2} $ = 1.953(4) d, respectively. P$ _{2} $ is approximately half of P$ _{1} $ and the disk precession cycle ($ \rm P_{1} \approx P _{prec}  \approx 2 \times P _{2} $). The variation with a period of P$ _{1} $ may come from the brightness change of the brightness center due to the accretion disk precession. The periodic variation of P$ _{2} $ is a new phenomenon found for the first time. A comparison of the frequency-amplitude plots for different sectors suggests that the variations in the eclipse depth are related to the accretion disk precession and are not necessarily linked to the white dwarf spin. Periodic variations in eclipse depth similar to P$ _{1} $ are found in NLs AQ Men, DW UMa, and SDSS J081256.85+191157.8, but there are no periodical variations similar to P$ _{2} $. Therefore, we suggest that the periodic variation of the eclipse depth with a period of P$ _{2} $ (about half of the accretion disk precession cycle) in TV Col is related to the uniqueness of the IP and the tilted accretion disk precession, possibly from the two accretion curtains precession together with the tilted disk. 

It is worth warning that our suggestions may be more optimistic and more testing and discussion is necessary. The complex variations observed in TV Col have important research value. Particularly noteworthy is the discovery of periodic variations in eclipse depth with a period of P$ _{2} $ = 1.953(4) d, which will provide crucial observational evidence for studying tilted disk precession and IPs.

\section*{Acknowledgements}

This work was supported by National Key R\&D Program of China (grant No. 2022YFE0116800), the National Natural Science Foundation of China (Nos. 11933008, 12103084, and 12303040) and the basic research project of Yunnan Province (Grant No. 202301AT070352). In this paper, we use is data from the TESS survey and thank all the staff who work with the TESS mission.
All of the data presented in this article were obtained from the Mikulski Archive for Space Telescopes (MAST) at the Space Telescope Science Institute. The specific observations analyzed can be accessed via \dataset[doi:10.17909/cdzh-vn81]{https://doi.org/10.17909/cdzh-vn81}.

\bibliography{sample631}{}

\begin{thebibliography}{}
\expandafter\ifx\csname natexlab\endcsname\relax\def\natexlab#1{#1}\fi
\providecommand{\url}[1]{\href{#1}{#1}}
\providecommand{\dodoi}[1]{doi:~\href{http://doi.org/#1}{\nolinkurl{#1}}}
\providecommand{\doeprint}[1]{\href{http://ascl.net/#1}{\nolinkurl{http://ascl.net/#1}}}
\providecommand{\doarXiv}[1]{\href{https://arxiv.org/abs/#1}{\nolinkurl{https://arxiv.org/abs/#1}}}

\bibitem[{{Armstrong} {et~al.}(2013){Armstrong}, {Patterson}, {Michelsen},
  {Thorstensen}, {Uthas}, {Vanmunster}, {Hambsch}, {Roberts}, \&
  {Dvorak}}]{2013MNRAS.435..707A}
{Armstrong}, E., {Patterson}, J., {Michelsen}, E., {et~al.} 2013, \mnras, 435,
  707, \dodoi{10.1093/mnras/stt1335}

\bibitem[{{Augusteijn} {et~al.}(1994){Augusteijn}, {Heemskerk}, {Zwarthoed}, \&
  {van Paradijs}}]{Augusteijn1994}
{Augusteijn}, T., {Heemskerk}, M.~H.~M., {Zwarthoed}, G.~A.~A., \& {van
  Paradijs}, J. 1994, \aaps, 107, 219

\bibitem[{{Barrett} {et~al.}(1988){Barrett}, {O'Donoghue}, \&
  {Warner}}]{Barrett1988MNRAS}
{Barrett}, P., {O'Donoghue}, D., \& {Warner}, B. 1988, \mnras, 233, 759,
  \dodoi{10.1093/mnras/233.4.759}

\bibitem[{{Bonnet-Bidaud} {et~al.}(1985){Bonnet-Bidaud}, {Motch}, \&
  {Mouchet}}]{1985A&A...143..313B}
{Bonnet-Bidaud}, J.~M., {Motch}, C., \& {Mouchet}, M. 1985, \aap, 143, 313

\bibitem[{{Boyd} {et~al.}(2017){Boyd}, {de Miguel}, {Patterson}, {Wood},
  {Barrett}, {Boardman}, {Brettman}, {Cejudo}, {Collins}, {Cook}, {Cook},
  {Foote}, {Fried}, {Gomez}, {Hambsch}, {Jones}, {Kemp}, {Koff}, {Koppelman},
  {Krajci}, {Lemay}, {Martin}, {McClusky}, {Menzies}, {Messier}, {Roberts},
  {Robertson}, {Rock}, {Sabo}, {Skillman}, {Ulowetz}, \&
  {Vanmunster}}]{Boyd2017}
{Boyd}, D.~R.~S., {de Miguel}, E., {Patterson}, J., {et~al.} 2017, \mnras, 466,
  3417, \dodoi{10.1093/mnras/stw3327}

\bibitem[{{Bruch}(2022)}]{Bruch2022}
{Bruch}, A. 2022, \mnras, 514, 4718, \dodoi{10.1093/mnras/stac1650}

\bibitem[{{Bruch}(2023{\natexlab{a}})}]{Bruch2023}
---. 2023{\natexlab{a}}, \mnras, 519, 352, \dodoi{10.1093/mnras/stac3493}

\bibitem[{{Bruch}(2023{\natexlab{b}})}]{BruchIII}
---. 2023{\natexlab{b}}, \mnras, \dodoi{10.1093/mnras/stad2089}

\bibitem[{{Charles} {et~al.}(1979){Charles}, {Thorstensen}, {Bowyer}, \&
  {Middleditch}}]{Charles1979ApJ}
{Charles}, P., {Thorstensen}, J., {Bowyer}, S., \& {Middleditch}, J. 1979,
  \apjl, 231, L131, \dodoi{10.1086/183019}

\bibitem[{{Crawford} {et~al.}(2008){Crawford}, {Boyd}, {Gualdoni}, {Gomez},
  {MacDonald}, \& {Oksanen}}]{2008JAVSO..36...60C}
{Crawford}, T., {Boyd}, D., {Gualdoni}, C., {et~al.} 2008, \jaavso, 36, 60

\bibitem[{{Cropper}(1990)}]{1990SSRv...54..195C}
{Cropper}, M. 1990, \ssr, 54, 195, \dodoi{10.1007/BF00177799}

\bibitem[{De~Miguel {et~al.}(2016)De~Miguel, Patterson, Cejudo, Ulowetz, Jones,
  Boardman, Barret, Koff, Stein, Campbell, {et~al.}}]{Miguel2016accretion}
De~Miguel, E., Patterson, J., Cejudo, D., {et~al.} 2016, \mnras, 457, 1447,
  \dodoi{10.1093/mnras/stv3014}

\bibitem[{Dubus {et~al.}(2018)Dubus, Otulakowska-Hypka, \&
  Lasota}]{dubus2018testing}
Dubus, G., Otulakowska-Hypka, M., \& Lasota, J.-P. 2018, \aap, 617, A26,
  \dodoi{10.1051/0004-6361/201833372}

\bibitem[{{Ezuka} \& {Ishida}(1999)}]{Ezuka1999}
{Ezuka}, H., \& {Ishida}, M. 1999, \apjs, 120, 277, \dodoi{10.1086/313181}

\bibitem[{{Ferrario}(1996)}]{1996PASA...13...87F}
{Ferrario}, L. 1996, \pasa, 13, 87, \dodoi{10.1017/S1323358000020592}

\bibitem[{{Hameury}(2020)}]{2020AdSpR..66.1004H}
{Hameury}, J.~M. 2020, Advances in Space Research, 66, 1004,
  \dodoi{10.1016/j.asr.2019.10.022}

\bibitem[{{Hameury} \& {Lasota}(2017)}]{2017A&A...602A.102H}
{Hameury}, J.~M., \& {Lasota}, J.~P. 2017, \aap, 602, A102,
  \dodoi{10.1051/0004-6361/201730760}

\bibitem[{Harvey {et~al.}(1995)Harvey, Skillman, Patterson, \&
  Ringwald}]{harvey1995superhumps}
Harvey, D., Skillman, D.~R., Patterson, J., \& Ringwald, F. 1995, \pasp, 107,
  551, \dodoi{10.1086/133591}

\bibitem[{Heil \& Walnut(1989)}]{CWT1989}
Heil, C.~E., \& Walnut, D.~F. 1989, SIAM Review, 31, 628,
  \dodoi{10.1137/1031129}

\bibitem[{{Hellier}(1993)}]{1993MNRAS.264..132H}
{Hellier}, C. 1993, \mnras, 264, 132, \dodoi{10.1093/mnras/264.1.132}

\bibitem[{{Hellier}(1995)}]{1995ASPC...85..185H}
{Hellier}, C. 1995, in Astronomical Society of the Pacific Conference Series,
  Vol.~85, Magnetic Cataclysmic Variables, ed. D.~A.~H. {Buckley} \&
  B.~{Warner}, 185

\bibitem[{{Hellier}(1999)}]{1999ApJ...519..324H}
---. 1999, \apj, 519, 324, \dodoi{10.1086/307345}

\bibitem[{{Hellier} {et~al.}(1991{\natexlab{a}}){Hellier}, {Cropper}, \&
  {Mason}}]{1991MNRAS.248..233H}
{Hellier}, C., {Cropper}, M., \& {Mason}, K.~O. 1991{\natexlab{a}}, \mnras,
  248, 233, \dodoi{10.1093/mnras/248.2.233}

\bibitem[{{Hellier} {et~al.}(2000){Hellier}, {Kemp}, {Naylor}, {Bateson},
  {Jones}, {Overbeek}, {Stubbings}, \& {Mukai}}]{2000MNRAS.313..703H}
{Hellier}, C., {Kemp}, J., {Naylor}, T., {et~al.} 2000, \mnras, 313, 703,
  \dodoi{10.1046/j.1365-8711.2000.03301.x}

\bibitem[{{Hellier} {et~al.}(1991{\natexlab{b}}){Hellier}, {Mason}, \&
  {Mittaz}}]{Hellier1991MNRAS}
{Hellier}, C., {Mason}, K.~O., \& {Mittaz}, J.~P.~D. 1991{\natexlab{b}},
  \mnras, 248, 5P, \dodoi{10.1093/mnras/248.1.5P}

\bibitem[{{Honeycutt} {et~al.}(1998){Honeycutt}, {Robertson}, {Turner}, \&
  {Mattei}}]{Honeycutt1998PASP}
{Honeycutt}, R.~K., {Robertson}, J.~W., {Turner}, G.~W., \& {Mattei}, J.~A.
  1998, \pasp, 110, 676, \dodoi{10.1086/316180}

\bibitem[{{Hudec} {et~al.}(2005){Hudec}, {{\v{S}}imon}, \&
  {Skalick{\'y}}}]{2005ASPC..330..405H}
{Hudec}, R., {{\v{S}}imon}, V., \& {Skalick{\'y}}, J. 2005, in Astronomical
  Society of the Pacific Conference Series, Vol. 330, The Astrophysics of
  Cataclysmic Variables and Related Objects, ed. J.~M. {Hameury} \& J.~P.
  {Lasota}, 405

\bibitem[{{Hutchings} {et~al.}(1981){Hutchings}, {Crampton}, {Cowley},
  {Thorstensen}, \& {Charles}}]{Hutchings1981ApJ}
{Hutchings}, J.~B., {Crampton}, D., {Cowley}, A.~P., {Thorstensen}, J.~R., \&
  {Charles}, P.~A. 1981, \apj, 249, 680, \dodoi{10.1086/159329}

\bibitem[{{I{\l}kiewicz} {et~al.}(2021){I{\l}kiewicz}, {Scaringi}, {Court},
  {Maccarone}, {Altamirano}, {Bradshaw}, {Degenaar}, {Fratta}, {Littlefield},
  {Shahbaz}, \& {Wijnands}}]{2021MNRAS.503.4050I}
{I{\l}kiewicz}, K., {Scaringi}, S., {Court}, J. M.~C., {et~al.} 2021, \mnras,
  503, 4050, \dodoi{10.1093/mnras/stab664}

\bibitem[{{Ishida} \& {Fujimoto}(1995)}]{Ishida1995}
{Ishida}, M., \& {Fujimoto}, R. 1995, in Astrophysics and Space Science
  Library, Vol. 205, Cataclysmic Variables, ed. A.~{Bianchini}, M.~{della
  Valle}, \& M.~{Orio}, 93, \dodoi{10.1007/978-94-011-0335-0_11}

\bibitem[{{Katz}(1973)}]{Katz1973NPhS..246...87K}
{Katz}, J.~I. 1973, Nature Physical Science, 246, 87,
  \dodoi{10.1038/physci246087a0}

\bibitem[{Lasota(2001)}]{lasota2001disc}
Lasota, J.-P. 2001, New Astronomy Reviews, 45, 449,
  \dodoi{10.1016/S1387-6473(01)00112-9}

\bibitem[{{Lenz} \& {Breger}(2005)}]{Period04}
{Lenz}, P., \& {Breger}, M. 2005, Communications in Asteroseismology, 146, 53,
  \dodoi{10.1553/cia146s53}

\bibitem[{{Lubow}(1991)}]{lubow1991model}
{Lubow}, S.~H. 1991, \apj, 381, 259, \dodoi{10.1086/170647}

\bibitem[{{Marino} \& {Walker}(1974)}]{1974IBVS..864....1M}
{Marino}, B.~F., \& {Walker}, W.~S.~G. 1974, Information Bulletin on Variable
  Stars, 864, 1

\bibitem[{{Meyer} \& {Meyer-Hofmeister}(1983)}]{Meyer1983A&A}
{Meyer}, F., \& {Meyer-Hofmeister}, E. 1983, \aap, 121, 29

\bibitem[{{Montgomery}(2009)}]{Montgomery2009MNRAS}
{Montgomery}, M.~M. 2009, \mnras, 394, 1897,
  \dodoi{10.1111/j.1365-2966.2009.14487.x}

\bibitem[{{Montgomery}(2012)}]{Montgomery2012}
---. 2012, Astrophysical Journal Letters, 745, L25,
  \dodoi{10.1088/2041-8205/745/2/L25}

\bibitem[{{Motch}(1981)}]{Motch1981}
{Motch}, C. 1981, \aap, 100, 277

\bibitem[{{Norton} {et~al.}(1999){Norton}, {Beardmore}, {Allan}, \&
  {Hellier}}]{1999A&A...347..203N}
{Norton}, A.~J., {Beardmore}, A.~P., {Allan}, A., \& {Hellier}, C. 1999, \aap,
  347, 203, \dodoi{10.48550/arXiv.astro-ph/9811310}

\bibitem[{Osaki(1985)}]{osaki1985irradiation}
Osaki, Y. 1985, \aap, 144, 369

\bibitem[{{Osaki}(1996)}]{1996PASP..108...39O}
{Osaki}, Y. 1996, \pasp, 108, 39, \dodoi{10.1086/133689}

\bibitem[{{Papaloizou} \& {Pringle}(1979)}]{1979MNRAS.189..293P}
{Papaloizou}, J., \& {Pringle}, J.~E. 1979, \mnras, 189, 293,
  \dodoi{10.1093/mnras/189.2.293}

\bibitem[{Patterson(1999)}]{patterson1999permanent}
Patterson, J. 1999, FRONTIERS SCIENCE SERIES, 61

\bibitem[{{Patterson} {et~al.}(1997){Patterson}, {Kemp}, {Saad}, {Skillman},
  {Harvey}, {Fried}, {Thorstensen}, \& {Ashley}}]{Patterson1997PASP}
{Patterson}, J., {Kemp}, J., {Saad}, J., {et~al.} 1997, \pasp, 109, 468,
  \dodoi{10.1086/133903}

\bibitem[{{Rana} {et~al.}(2004){Rana}, {Singh}, {Schlegel}, \&
  {Barrett}}]{Rana2004}
{Rana}, V.~R., {Singh}, K.~P., {Schlegel}, E.~M., \& {Barrett}, P. 2004, \aj,
  127, 489, \dodoi{10.1086/380232}

\bibitem[{{Retter} {et~al.}(2003){Retter}, {Hellier}, {Augusteijn}, {Naylor},
  {Bedding}, {Bembrick}, {McCormick}, \& {Velthuis}}]{2003MNRAS.340..679R}
{Retter}, A., {Hellier}, C., {Augusteijn}, T., {et~al.} 2003, \mnras, 340, 679,
  \dodoi{10.1046/j.1365-8711.2003.06331.x}

\bibitem[{Ricker {et~al.}(2015)Ricker, Winn, \& Vanderspek}]{Ricker2015journal}
Ricker, G., Winn, J., \& Vanderspek, R. 2015, JATIS, 1, 014003,
  \dodoi{10.1117/1.JATIS.1.1.014003}

\bibitem[{{Scaringi} {et~al.}(2022){Scaringi}, {Groot}, {Knigge}, {Bird},
  {Breedt}, {Buckley}, {Cavecchi}, {Degenaar}, {de Martino}, {Done}, {Fratta},
  {I{\l}kiewicz}, {Koerding}, {Lasota}, {Littlefield}, {Manara}, {O'Brien},
  {Szkody}, \& {Timmes}}]{Scaringi2022Natur}
{Scaringi}, S., {Groot}, P.~J., {Knigge}, C., {et~al.} 2022, \nat, 604, 447,
  \dodoi{10.1038/s41586-022-04495-6}

\bibitem[{{Schrijver} {et~al.}(1987){Schrijver}, {Brinkman}, \& {van der
  Woerd}}]{Schrijver1987}
{Schrijver}, J., {Brinkman}, A.~C., \& {van der Woerd}, H. 1987, \apss, 130,
  261, \dodoi{10.1007/BF00655004}

\bibitem[{{Schwarz} \& {Heemskerk}(1987)}]{1987IAUC.4508....1S}
{Schwarz}, H.~E., \& {Heemskerk}, M.~H.~M. 1987, \iaucirc, 4508, 1

\bibitem[{{Shakura} \& {Sunyaev}(1973)}]{Shakura1973}
{Shakura}, N.~I., \& {Sunyaev}, R.~A. 1973, \aap, 24, 337

\bibitem[{{Smak}(1983)}]{Smak1983ApJ}
{Smak}, J. 1983, \apj, 272, 234, \dodoi{10.1086/161284}

\bibitem[{{Stefanov} \& {Stefanov}(2023)}]{2023MNRAS.520.3355S}
{Stefanov}, S.~Y., \& {Stefanov}, A.~K. 2023, \mnras, 520, 3355,
  \dodoi{10.1093/mnras/stad259}

\bibitem[{Sun {et~al.}(2023)Sun, Qian, \& Li}]{2023arXiv230905891S}
Sun, Q.-B., Qian, S.-B., \& Li, M.-Y. 2023, \apj, 955, 135,
  \dodoi{10.3847/1538-4357/ace183}

\bibitem[{{Sun} {et~al.}(2023{\natexlab{a}}){Sun}, {Qian}, {Zhu}, {Liao},
  {Zhao}, {Li}, {Shi}, \& {Li}}]{Sun2023arXiv}
{Sun}, Q.-B., {Qian}, S.-B., {Zhu}, L.-Y., {et~al.} 2023{\natexlab{a}}, \mnras,
  526, 3730, \dodoi{10.1093/mnras/stad1880}

\bibitem[{{Sun} {et~al.}(2024){Sun}, {Qian}, {Zhu}, {Liao}, {Zhao}, {Li},
  {Shi}, \& {Li}}]{2023arXiv230911033S}
---. 2024, \apj, 962, 123, \dodoi{10.3847/1538-4357/ad0f1c}

\bibitem[{{Sun} {et~al.}(2022){Sun}, {Qian}, {Dong}, {Zhi}, {Han}, {Liu},
  {Chang}, {Liu}, {Xiang}, {Peng}, {Zhang}, {Zhang}, \& {Fern{\'a}ndez
  Laj{\'u}s}}]{sun2022study}
{Sun}, Q.-B., {Qian}, S.-B., {Dong}, A.-J., {et~al.} 2022, \na, 93, 101751,
  \dodoi{10.1016/j.newast.2021.101751}

\bibitem[{{Sun} {et~al.}(2023{\natexlab{b}}){Sun}, {Qian}, {Zhu}, {Dong},
  {Zhi}, {Liao}, {Zhao}, {Han}, {Liu}, {Zang}, {Li}, \& {Shi}}]{Sun2023MNRAS}
{Sun}, Q.-B., {Qian}, S.-B., {Zhu}, L.-Y., {et~al.} 2023{\natexlab{b}}, \mnras,
  518, 3901, \dodoi{10.1093/mnras/stac3272}

\bibitem[{{Szkody} \& {Mateo}(1984)}]{Szkody1984ApJ}
{Szkody}, P., \& {Mateo}, M. 1984, \apj, 280, 729, \dodoi{10.1086/162045}

\bibitem[{{Vogt}(1974)}]{1974A&A....36..369V}
{Vogt}, N. 1974, \aap, 36, 369

\bibitem[{Vogt(1982)}]{vogt1982z}
Vogt, N. 1982, \apj, 252, 653, \dodoi{10.1086/159592}

\bibitem[{{Vrtilek} {et~al.}(1996){Vrtilek}, {Silber}, {Primini}, \&
  {Raymond}}]{Vrtilek1996}
{Vrtilek}, S.~D., {Silber}, A., {Primini}, F., \& {Raymond}, J.~C. 1996, \apj,
  465, 951, \dodoi{10.1086/177479}

\bibitem[{{Warner}(1975)}]{Warner1975MNRAS.170..219W}
{Warner}, B. 1975, \mnras, 170, 219, \dodoi{10.1093/mnras/170.1.219}

\bibitem[{{Warner}(1985)}]{1985ASIC..150..367W}
{Warner}, B. 1985, in NATO Advanced Study Institute (ASI) Series C, Vol. 150,
  Interacting Binaries, ed. P.~P. {Eggleton} \& J.~E. {Pringle}, 367,
  \dodoi{10.1007/978-94-009-5337-6_14}

\bibitem[{Warner(1995)}]{warner1995cat}
Warner, B. 1995, Cambridge University Press, 28

\bibitem[{{Whitehurst}(1988)}]{1988MNRAS.232...35W}
{Whitehurst}, R. 1988, \mnras, 232, 35, \dodoi{10.1093/mnras/232.1.35}

\bibitem[{{Wood} \& {Burke}(2007)}]{Wood2007ApJ...661.1042W}
{Wood}, M.~A., \& {Burke}, C.~J. 2007, Astrophysical Journal, 661, 1042,
  \dodoi{10.1086/516723}

\bibitem[{{Wood} {et~al.}(2000){Wood}, {Montgomery}, \&
  {Simpson}}]{Wood2000ApJ}
{Wood}, M.~A., {Montgomery}, M.~M., \& {Simpson}, J.~C. 2000, Astrophysical
  Journal Letters, 535, L39, \dodoi{10.1086/312687}

\bibitem[{Wood {et~al.}(2011)Wood, Still, Howell, Cannizzo, \&
  Smale}]{wood2011v344}
Wood, M.~A., Still, M.~D., Howell, S.~B., Cannizzo, J.~K., \& Smale, A.~P.
  2011, \apj, 741, 105, \dodoi{10.1088/0004-637X/741/2/105}

\bibitem[{Wood {et~al.}(2009)Wood, Thomas, \& Simpson}]{wood2009sph}
Wood, M.~A., Thomas, D.~M., \& Simpson, J.~C. 2009, \mnras, 398, 2110,
  \dodoi{10.1111/j.1365-2966.2009.15252.x}

\end{thebibliography}
\bibliographystyle{aasjournal}

\section{Appendix}\label{sec:Appendix}
Appendix contains Tables \ref{tab:A1} and \ref{tab:A2}. Table \ref{tab:A1} shows the segmented superposition sinusoidal fit parameters, as detailed in Section \ref{subsec:amplitude}. Table \ref{tab:A2} shows the eclipse parameters, including the eclipse depth and O-C analysis.

\renewcommand\thetable{A1}
\setlength{\tabcolsep}{1pt}
\begin{longtable}{ccccccccccccccccc}

	\caption{Results of the segmented fit to the NSH using linear superimposed sine fit.}\label{tab:A1}\\

	
	\toprule
	Start&End&during&Z&err&B&errors&A&errors&freq&errors&p&errors&$\rm \nu$&$\rm \chi_{red}^2$\\
	
	[BJD-2457000]&[BJD-2457000]&[d]&[e/s]&[e/s]&[e/(s$ \cdot $d)]&[e/(s$ \cdot $d)]&[e/s]&[e/s]&[1/d]&[1/d]&[rad] &[rad] &&\\
	\midrule
	\endfirsthead
	\multicolumn{15}{c}
	{{\bfseries{Table \thetable\ }continued from previous page}} \\
	\toprule
	Start&End&during&Z&err&B&errors&A&errors&freq&errors&p&errors&$\rm \nu$&$\rm \chi_{red}^2$\\

[BJD-2457000]&[BJD-2457000]&[d]&[e/s]&[e/s]&[e/(s$ \cdot $d)]&[e/(s$ \cdot $d)]&[e/s]&[e/s]&[1/d]&[1/d]&[rad] &[rad] &&\\
	\midrule
	\endhead
	\midrule
	\multicolumn{15}{c}
	{{ }} \\
	\endfoot
	\endlastfoot
2174.2581&2175.2553&0.9972&157831.2533&9878.923&-72.564&4.543&-19.730&6.371&4.992&0.047&733.964&102.270&555&1.576 \\
2175.2581&2176.1720&0.9139&-158839.8573&10821.198&73.004&4.974&46.513&7.602&4.673&0.025&-85.146&55.271&496&1.682 \\
2176.1748&2177.0887&0.9139&35247.9350&10333.078&-16.182&4.747&-65.998&6.790&4.531&0.016&235.518&35.247&496&1.224 \\
2177.0915&2178.0054&0.9139&-1068.0941&9166.695&0.510&4.210&73.049&6.337&4.598&0.013&101.231&29.092&497&1.032 \\
2178.0082&2178.9221&0.9139&210745.2648&14014.450&-96.745&6.433&42.735&9.233&4.993&0.031&-785.536&67.987&496&2.050 \\
2178.9249&2179.8388&0.9139&71705.0663&10797.875&-32.925&4.955&56.894&6.691&4.462&0.019&406.947&41.445&496&1.712 \\
2179.8415&2180.7554&0.9139&-62777.4626&7689.778&28.769&3.527&-66.486&5.002&4.613&0.012&55.724&25.104&497&0.701 \\
2180.7582&2181.6721&0.9139&-98324.3357&11070.945&45.068&5.076&83.653&8.000&4.603&0.014&45.311&31.234&496&1.285 \\
2181.6749&2182.5888&0.9139&-88882.5663&10061.945&40.728&4.611&50.184&6.708&4.726&0.022&459.499&47.144&496&1.435 \\
2182.5916&2183.5055&0.9139&63994.2584&10314.140&-29.321&4.725&52.846&5.881&4.382&0.019&956.716&40.736&496&1.523 \\
2183.5082&2184.4221&0.9139&-16480.7081&10848.408&7.533&4.967&58.933&6.667&4.555&0.018&176.446&38.604&495&1.518 \\
2184.4249&2185.3388&0.9139&-138771.4989&15978.044&63.498&7.313&-85.979&10.440&4.496&0.019&299.217&41.745&496&2.747 \\
2185.3416&2185.7805&0.4389&286448.1342&42169.455&-131.033&19.295&65.435&22.578&4.843&0.078&-591.709&169.945&233&5.072 \\
2187.1944&2188.1083&0.9139&177977.7125&19830.892&-81.331&9.065&-65.674&11.549&4.425&0.028&559.140&60.571&495&5.372 \\
2188.1110&2189.0249&0.9139&43989.2942&9630.846&-20.096&4.401&65.357&7.095&4.531&0.017&276.674&36.969&495&1.009 \\
2189.0277&2189.9416&0.9139&55222.1583&11371.735&-25.217&5.194&-61.728&8.093&4.781&0.020&-335.418&44.597&495&1.818 \\
2189.9444&2190.8583&0.9139&18133.7084&10705.283&-8.291&4.887&25.659&8.162&4.760&0.045&-274.233&98.126&495&1.451 \\
2190.8611&2191.7750&0.9139&6195.2703&11256.903&-2.832&5.137&39.120&6.329&4.347&0.027&1223.470&58.456&496&1.800 \\
2191.7777&2192.6916&0.9139&11158.5577&7449.735&-5.102&3.398&60.886&5.195&4.621&0.013&29.904&27.919&496&0.655 \\
2192.6944&2193.6083&0.9139&-214841.9781&12981.948&97.971&5.919&79.237&9.579&4.611&0.018&85.578&39.898&488&2.290 \\
2193.6111&2194.5250&0.9139&-56413.4726&9610.983&25.720&4.380&26.500&5.505&4.713&0.037&1667.449&80.773&496&1.147 \\
2194.5277&2195.6902&1.1625&17969.5543&5598.501&-8.187&2.550&31.739&3.990&4.690&0.019&-127.961&42.561&634&0.701 \\
2202.8332&2203.7471&0.9139&107836.9250&7743.623&-48.930&3.515&25.305&4.976&4.606&0.031&57.491&68.603&496&0.785 \\
2203.7499&2204.6638&0.9139&-3032.9058&9560.046&1.381&4.337&-28.698&6.904&4.605&0.035&56.034&76.702&497&1.173 \\
2204.6666&2205.5804&0.9139&-166084.1419&8554.024&75.331&3.879&27.975&5.430&4.595&0.031&76.923&67.337&496&0.957 \\
2205.5832&2206.4971&0.9139&227303.3450&10586.992&-103.042&4.799&41.918&7.207&4.575&0.026&199.672&58.015&497&1.428 \\
2206.4999&2207.4137&0.9139&66019.0171&9956.680&-29.937&4.512&55.846&5.766&4.464&0.016&376.957&36.361&496&1.356 \\
2207.4165&2208.3304&0.9139&11528.2198&8615.483&-5.245&3.902&-57.569&6.151&4.660&0.016&-61.282&34.499&489&0.997 \\
2208.3332&2209.2470&0.9139&-149736.9449&11071.987&67.777&5.013&66.170&7.659&4.630&0.017&5.888&37.994&497&1.536 \\
2209.2498&2210.1637&0.9139&-36066.1157&8645.562&16.314&3.913&-29.882&6.051&4.849&0.029&-479.700&63.747&496&0.986 \\
2210.1665&2211.0803&0.9139&-45558.8121&11041.256&20.605&4.995&-53.119&6.479&4.494&0.019&308.318&42.270&497&1.629 \\
2211.0831&2211.9970&0.9139&-27637.8244&9919.021&12.502&4.485&59.719&6.824&4.490&0.018&319.179&39.055&496&1.309 \\
2211.9998&2212.9136&0.9139&-125174.9683&11047.086&56.595&4.993&-82.738&8.309&4.640&0.014&-16.123&31.836&496&1.443 \\
2212.9164&2213.8664&0.9500&78210.9765&10711.356&-35.326&4.839&34.712&7.724&4.815&0.034&-403.738&74.517&522&1.704 \\
2215.4386&2216.3525&0.9139&-91571.1856&9143.389&41.332&4.126&80.693&7.188&4.544&0.013&243.744&29.763&496&1.023 \\
2216.3552&2217.2691&0.9139&20007.7376&11642.403&-9.011&5.252&-75.474&7.110&4.548&0.014&199.815&31.892&496&1.645 \\
2217.2719&2218.1858&0.9139&130111.9996&12035.440&-58.670&5.427&31.952&8.397&4.853&0.037&-488.361&81.072&495&1.663 \\
2218.1885&2219.1024&0.9139&102393.9553&10310.364&-46.163&4.647&57.235&6.514&4.591&0.017&105.511&38.595&496&1.486 \\
2219.1052&2220.0190&0.9139&-65354.1610&8508.033&29.435&3.833&-77.825&6.198&4.524&0.012&308.677&27.183&495&0.952 \\
2220.0218&2220.9357&0.9139&-118065.1114&12297.070&53.172&5.538&-90.640&8.064&4.631&0.013&24.165&28.992&496&1.648 \\
2220.9385&2221.8523&0.9139&86568.6754&12181.249&-38.974&5.484&31.509&9.500&4.700&0.044&-112.927&97.209&489&2.068 \\
2221.8551&2222.7690&0.9139&85583.6566&11448.357&-38.528&5.152&-49.187&6.959&4.584&0.022&145.086&48.050&494&1.764 \\
2222.7717&2223.6856&0.9139&153806.0119&12808.332&-69.193&5.761&-56.829&8.826&4.437&0.026&703.819&56.893&497&1.923 \\
2223.6884&2224.6022&0.9139&3667.0583&12100.076&-1.656&5.440&82.184&7.536&4.628&0.013&-0.383&28.748&496&1.572 \\
2224.6050&2225.5189&0.9139&-112491.7545&12250.916&50.568&5.506&-51.544&8.098&4.638&0.024&3.854&52.751&496&2.087 \\
2225.5217&2226.4355&0.9139&31951.3553&10823.902&-14.350&4.862&44.663&6.868&4.677&0.023&-54.317&51.321&496&1.381 \\
2226.4383&2227.5744&1.1361&-53667.3574&7043.136&24.102&3.163&63.567&5.306&4.601&0.012&82.326&27.294&616&1.031 \\
	\bottomrule
\end{longtable}

\renewcommand\thetable{A2}
\setlength{\tabcolsep}{0.1pt}
\begin{longtable}{ccccccccccccccccc}

	\caption{Eclipse parameters and O-C analysis .}\label{tab:A2}\\

	
	\toprule
	Minima-BJD&err&Minima-flux&err&Eclipse depth&E&O-C&Minima-BJD&err&Minima-flux&err&Eclipse depth&E&O-C\\

[BJD - 2457000]&[d]&[e/s]&[e/s]&[e/s]& &[d]&[BJD - 2457000]&[d]&[e/s]&[e/s]&[e/s]& &[d]\\
	\midrule
	\endfirsthead
	\multicolumn{14}{c}
	{{\bfseries{Table \thetable\ }continued from previous page}} \\
	\toprule
	Minima-BJD&err&Minima-flux&err&Eclipse depth&E&O-C&Minima-BJD&err&Minima-flux&err&Eclipse depth&E&O-C\\
	
	[BJD - 2457000]&[d]&[e/s]&[e/s]&[e/s]& &[d]&[BJD - 2457000]&[d]&[e/s]&[e/s]&[e/s]& &[d]\\
	\midrule
	\endhead
	\midrule
	\multicolumn{14}{c}
	{{ }} \\
	\endfoot
	\endlastfoot
2174.45338 &0.00060 &-92.853 &13.966 &86.711 &0&0.00000 &2204.40009 &0.00045 &-119.439 &13.797 &122.630 &131&0.00009 \\
2174.67992 &0.00044 &-57.901 &8.901 &52.181 &1&-0.00206 &2204.63025 &0.00037 &-134.902 &11.374 &123.832 &132&0.00165 \\
2174.91350 &0.00115 &-29.619 &11.753 &18.341 &2&0.00291 &2204.86073 &0.00065 &-111.981 &15.685 &112.544 &133&0.00353 \\
2175.14597 &0.00096 &-25.227 &11.909 &29.096 &3&0.00679 &2205.09166 &0.00054 &-99.684 &13.405 &99.172 &134&0.00586 \\
2175.38107 &0.00091 &-16.827 &11.583 &28.753 &4&0.01329 &2205.31671 &0.00036 &-151.050 &13.596 &146.005 &135&0.00232 \\
2175.60374 &0.00074 &-51.161 &10.742 &48.576 &5&0.00736 &2205.54580 &0.00038 &-140.992 &9.201 &116.383 &136&0.00281 \\
2175.83123 &0.00053 &-120.847 &9.818 &97.053 &6&0.00625 &2205.77489 &0.00049 &-66.599 &11.259 &78.221 &137&0.00329 \\
2176.05993 &0.00038 &-107.178 &11.097 &123.232 &7&0.00635 &2206.45892 &0.00091 &-52.002 &14.044 &60.325 &140&0.00153 \\
2176.28522 &0.00056 &-97.048 &12.064 &101.243 &8&0.00304 &2206.69172 &0.00060 &-64.362 &13.737 &68.575 &141&0.00572 \\
2176.51559 &0.00076 &-100.831 &12.672 &82.876 &9&0.00481 &2206.91835 &0.00043 &-66.593 &9.812 &81.463 &142&0.00375 \\
2176.74133 &0.00061 &-82.123 &12.010 &77.449 &10&0.00194 &2207.14520 &0.00072 &-80.423 &11.263 &62.425 &143&0.00200 \\
2176.97350 &0.00046 &-86.319 &12.901 &80.445 &11&0.00552 &2207.38078 &0.00050 &-95.984 &9.169 &88.757 &144&0.00898 \\
2177.20367 &0.00075 &-113.397 &15.038 &92.714 &12&0.00709 &2207.60703 &0.00046 &-106.745 &8.677 &78.521 &145&0.00663 \\
2177.42858 &0.00068 &-129.634 &18.763 &120.066 &13&0.00340 &2207.83301 &0.00035 &-133.786 &9.478 &130.346 &146&0.00401 \\
2177.65740 &0.00040 &-146.683 &12.320 &145.235 &14&0.00361 &2208.06275 &0.00036 &-127.098 &11.782 &123.589 &147&0.00515 \\
2177.88345 &0.00038 &-155.562 &11.007 &126.276 &15&0.00107 &2208.28865 &0.00046 &-150.293 &17.262 &145.194 &148&0.00246 \\
2178.10992 &0.00069 &-129.005 &17.279 &120.877 &16&-0.00107 &2208.51713 &0.00047 &-138.140 &13.398 &121.494 &149&0.00234 \\
2178.34067 &0.00085 &-94.913 &17.084 &79.037 &17&0.00108 &2208.74537 &0.00077 &-115.353 &14.300 &115.021 &150&0.00197 \\
2178.56250 &0.00068 &-50.724 &13.854 &39.159 &18&-0.00568 &2208.97781 &0.00052 &-138.932 &12.681 &112.890 &151&0.00582 \\
2178.80649 &0.00098 &1.806 &10.531 &5.588 &19&0.00970 &2209.20191 &0.00044 &-183.045 &19.728 &158.089 &152&0.00131 \\
2179.02817 &0.00127 &2.477 &20.649 &15.468 &20&0.00279 &2209.42661 &0.00048 &-108.099 &19.001 &119.787 &153&-0.00259 \\
2179.25697 &0.00068 &-53.602 &12.422 &68.375 &21&0.00298 &2209.89163 &0.00103 &-35.806 &18.662 &33.694 &155&0.00523 \\
2179.48944 &0.00074 &-61.731 &15.676 &63.989 &22&0.00685 &2210.12062 &0.00069 &-57.261 &15.101 &55.594 &156&0.00562 \\
2179.71665 &0.00061 &-78.728 &9.433 &66.207 &23&0.00547 &2210.34901 &0.00067 &-30.995 &10.891 &47.992 &157&0.00542 \\
2179.94625 &0.00074 &-74.288 &11.493 &54.916 &24&0.00646 &2210.57868 &0.00051 &-65.802 &12.803 &78.861 &158&0.00648 \\
2180.17539 &0.00045 &-97.570 &9.095 &106.601 &25&0.00700 &2210.80109 &0.00071 &-87.021 &13.988 &82.640 &159&0.00029 \\
2180.40183 &0.00040 &-133.350 &11.092 &125.669 &26&0.00484 &2211.03299 &0.00048 &-109.348 &11.620 &110.552 &160&0.00359 \\
2180.62418 &0.00049 &-108.890 &15.794 &103.412 &27&-0.00140 &2211.26486 &0.00046 &-125.776 &11.648 &124.868 &161&0.00686 \\
2180.85783 &0.00071 &-68.094 &16.448 &59.095 &28&0.00365 &2211.49182 &0.00022 &-168.598 &8.707 &145.011 &162&0.00522 \\
2181.08900 &0.00099 &-84.967 &18.310 &79.257 &29&0.00622 &2211.71297 &0.00055 &-113.963 &13.928 &111.465 &163&-0.00223 \\
2181.31710 &0.00048 &-130.972 &12.796 &123.929 &30&0.00571 &2211.94735 &0.00054 &-101.236 &15.395 &99.736 &164&0.00355 \\
2181.54476 &0.00046 &-168.811 &16.163 &155.608 &31&0.00478 &2212.17544 &0.00075 &-112.341 &23.748 &91.895 &165&0.00304 \\
2181.76941 &0.00043 &-102.111 &13.847 &119.134 &32&0.00082 &2212.40535 &0.00069 &-130.951 &24.691 &121.155 &166&0.00436 \\
2181.99557 &0.00037 &-117.824 &13.071 &141.353 &33&-0.00162 &2212.63575 &0.00049 &-171.387 &16.988 &165.355 &167&0.00615 \\
2182.22624 &0.00126 &-74.044 &16.388 &71.864 &34&0.00045 &2212.86257 &0.00035 &-202.151 &15.170 &196.097 &168&0.00437 \\
2182.45426 &0.00060 &-102.064 &18.369 &77.502 &35&-0.00013 &2213.08456 &0.00057 &-91.286 &15.644 &121.437 &169&-0.00224 \\
2182.69337 &0.00108 &20.135 &14.399 &4.735 &36&0.01038 &2213.31242 &0.00069 &-97.496 &18.564 &95.020 &170&-0.00298 \\
2183.14609 &0.00092 &-30.658 &15.690 &36.805 &38&0.00591 &2213.54059 &0.00103 &-101.432 &26.007 &75.238 &171&-0.00341 \\
2183.37246 &0.00081 &-65.013 &15.743 &72.379 &39&0.00367 &2213.77407 &0.00092 &-45.834 &13.822 &44.268 &172&0.00147 \\
2183.60303 &0.00063 &-69.894 &12.862 &85.811 &40&0.00564 &2215.60285 &0.00044 &-103.462 &14.044 &104.195 &180&0.00145 \\
2183.83296 &0.00038 &-107.556 &8.174 &88.754 &41&0.00697 &2215.82851 &0.00089 &-79.459 &15.101 &80.707 &181&-0.00149 \\
2184.06020 &0.00052 &-88.895 &7.585 &79.023 &42&0.00561 &2216.06622 &0.00050 &-103.886 &15.811 &102.658 &182&0.00762 \\
2184.28700 &0.00049 &-131.002 &11.715 &104.163 &43&0.00381 &2216.29222 &0.00056 &-119.629 &19.621 &114.436 &183&0.00502 \\
2184.51550 &0.00031 &-84.498 &8.259 &92.354 &44&0.00371 &2216.52301 &0.00063 &-153.984 &19.982 &139.726 &184&0.00721 \\
2184.74573 &0.00048 &-68.121 &10.151 &74.387 &45&0.00534 &2216.74709 &0.00040 &-158.085 &14.779 &166.538 &185&0.00269 \\
2184.97718 &0.00077 &-89.909 &12.615 &67.412 &46&0.00819 &2216.97394 &0.00055 &-156.485 &19.638 &137.114 &186&0.00094 \\
2185.20413 &0.00048 &-126.102 &14.197 &121.479 &47&0.00654 &2217.20267 &0.00039 &-149.274 &13.334 &123.575 &187&0.00107 \\
2185.43026 &0.00035 &-166.641 &10.553 &136.821 &48&0.00407 &2217.43007 &0.00125 &-69.321 &19.906 &70.809 &188&-0.00013 \\
2185.65514 &0.00039 &-172.240 &13.116 &145.224 &49&0.00036 &2217.65865 &0.00044 &-64.210 &10.633 &62.156 &189&-0.00015 \\
2187.25541 &0.00080 &-165.210 &19.689 &103.694 &56&0.00042 &2217.88674 &0.00076 &-38.352 &11.975 &36.258 &190&-0.00066 \\
2187.48429 &0.00059 &-81.278 &13.188 &82.167 &57&0.00070 &2218.11893 &0.00104 &-55.809 &20.352 &47.503 &191&0.00293 \\
2187.71917 &0.00052 &-89.482 &13.562 &93.708 &58&0.00698 &2218.34953 &0.00060 &-83.855 &15.731 &74.760 &192&0.00493 \\
2187.94759 &0.00020 &-107.326 &5.974 &133.127 &59&0.00681 &2218.57521 &0.00092 &-76.236 &18.860 &75.740 &193&0.00201 \\
2188.17008 &0.00045 &-116.644 &15.344 &121.975 &60&0.00069 &2218.80613 &0.00040 &-109.390 &8.982 &106.616 &194&0.00433 \\
2188.40128 &0.00048 &-98.740 &13.414 &97.901 &61&0.00330 &2219.03397 &0.00038 &-108.557 &9.214 &114.140 &195&0.00356 \\
2188.63076 &0.00040 &-122.507 &11.113 &114.945 &62&0.00417 &2219.26424 &0.00053 &-108.283 &16.424 &123.405 &196&0.00524 \\
2188.85992 &0.00041 &-130.905 &10.727 &121.039 &63&0.00473 &2219.48980 &0.00064 &-100.660 &19.100 &108.308 &197&0.00220 \\
2189.08903 &0.00040 &-147.182 &9.431 &117.133 &64&0.00524 &2219.72139 &0.00103 &-76.963 &19.617 &61.745 &198&0.00519 \\
2189.31452 &0.00034 &-137.859 &10.204 &149.276 &65&0.00213 &2219.95088 &0.00056 &-78.495 &14.826 &78.763 &199&0.00607 \\
2189.54062 &0.00043 &-156.411 &12.478 &147.679 &66&-0.00037 &2220.18185 &0.00040 &-136.605 &11.389 &115.508 &200&0.00844 \\
2189.76864 &0.00050 &-142.125 &11.857 &116.470 &67&-0.00095 &2220.40569 &0.00040 &-137.932 &15.033 &165.995 &201&0.00369 \\
2190.22684 &0.00111 &10.126 &12.918 &2.949 &69&0.00006 &2220.63543 &0.00037 &-175.873 &11.956 &158.052 &202&0.00482 \\
2190.46012 &0.00112 &-11.192 &11.601 &-1.220 &70&0.00473 &2220.86035 &0.00037 &-175.527 &12.613 &131.379 &203&0.00115 \\
2190.68208 &0.00145 &-24.861 &20.643 &-1.561 &71&-0.00191 &2221.08666 &0.00061 &-75.402 &18.242 &104.470 &204&-0.00114 \\
2190.91901 &0.00124 &-76.729 &21.052 &121.621 &72&0.00642 &2221.31539 &0.00070 &-70.019 &20.493 &76.718 &205&-0.00101 \\
2191.14912 &0.00071 &-35.356 &9.949 &53.541 &73&0.00793 &2221.54336 &0.00071 &-73.699 &13.627 &56.582 &206&-0.00164 \\
2191.37529 &0.00045 &-69.345 &9.494 &53.473 &74&0.00550 &2221.77425 &0.00053 &-61.791 &10.482 &57.831 &207&0.00065 \\
2191.60203 &0.00071 &-84.783 &18.517 &72.906 &75&0.00364 &2222.00234 &0.00121 &-23.010 &20.326 &40.770 &208&0.00014 \\
2191.83017 &0.00163 &-59.582 &16.326 &85.323 &76&0.00318 &2222.23593 &0.00085 &-59.488 &12.695 &37.585 &209&0.00512 \\
2192.06207 &0.00041 &-100.498 &10.065 &90.495 &77&0.00648 &2222.46435 &0.00098 &-54.032 &15.704 &65.871 &210&0.00495 \\
2192.29196 &0.00087 &-83.746 &12.446 &75.095 &78&0.00777 &2222.69503 &0.00064 &-97.832 &11.627 &84.168 &211&0.00703 \\
2192.51388 &0.00035 &-108.340 &10.099 &111.498 &79&0.00109 &2222.92029 &0.00042 &-137.965 &10.498 &127.912 &212&0.00368 \\
2192.75011 &0.00052 &-77.766 &14.432 &67.437 &80&0.00872 &2223.15328 &0.00046 &-91.102 &11.024 &104.371 &213&0.00808 \\
2192.97554 &0.00050 &-122.434 &12.804 &135.954 &81&0.00555 &2223.37424 &0.00052 &-148.015 &15.777 &113.715 &214&0.00043 \\
2193.20192 &0.00030 &-185.799 &10.230 &141.076 &82&0.00333 &2223.60429 &0.00080 &-87.333 &18.277 &80.938 &215&0.00189 \\
2193.42777 &0.00050 &-167.653 &13.258 &179.259 &83&0.00058 &2223.83580 &0.00076 &-71.326 &15.470 &59.873 &216&0.00480 \\
2193.65369 &0.00054 &-64.829 &13.258 &86.933 &84&-0.00210 &2224.06374 &0.00067 &-110.847 &18.647 &117.192 &217&0.00413 \\
2193.88707 &0.00107 &-25.715 &13.152 &-5.335 &85&0.00268 &2224.29593 &0.00050 &-147.609 &14.843 &125.194 &218&0.00772 \\
2194.11727 &0.00047 &-100.891 &8.497 &112.869 &86&0.00428 &2224.52045 &0.00057 &-123.665 &15.799 &126.903 &219&0.00364 \\
2194.34295 &0.00077 &-63.730 &14.994 &81.516 &87&0.00136 &2224.74579 &0.00047 &-138.555 &13.660 &136.449 &220&0.00038 \\
2194.57049 &0.00042 &-63.871 &9.777 &69.023 &88&0.00029 &2224.97267 &0.00038 &-108.215 &11.052 &134.292 &221&-0.00134 \\
2194.79828 &0.00046 &-74.115 &8.655 &54.776 &89&-0.00051 &2225.20520 &0.00060 &-108.654 &13.812 &89.164 &222&0.00259 \\
2195.02811 &0.00037 &-112.695 &7.070 &118.538 &90&0.00072 &2225.43183 &0.00110 &-92.286 &24.937 &61.409 &223&0.00062 \\
2195.25503 &0.00082 &-59.422 &12.252 &30.454 &91&-0.00097 &2225.66213 &0.00102 &-21.654 &18.522 &29.054 &224&0.00233 \\
2195.48592 &0.00042 &-96.866 &6.098 &87.098 &92&0.00132 &2225.89422 &0.00082 &-24.392 &12.872 &41.726 &225&0.00582 \\
2203.03168 &0.00043 &-98.441 &11.408 &103.110 &125&0.00329 &2226.34765 &0.00088 &-69.702 &18.011 &76.932 &227&0.00205 \\
2203.25897 &0.00062 &-103.722 &12.391 &100.876 &126&0.00198 &2226.57607 &0.00081 &-75.158 &16.760 &76.956 &228&0.00187 \\
2203.49130 &0.00047 &-109.830 &9.824 &99.173 &127&0.00571 &2226.80438 &0.00088 &-77.695 &18.765 &76.832 &229&0.00158 \\
2203.71902 &0.00074 &-75.807 &13.435 &69.201 &128&0.00482 &2227.03447 &0.00037 &-110.820 &10.906 &117.527 &230&0.00306 \\
2203.94786 &0.00041 &-108.752 &10.234 &110.775 &129&0.00507 &2227.26451 &0.00034 &-139.880 &11.001 &125.225 &231&0.00451 \\
2204.17518 &0.00044 &-139.497 &11.897 &107.638 &130&0.00378 &2227.48525 &0.00060 &-85.731 &13.196 &80.345 &232&-0.00336 \\
	\bottomrule
\end{longtable}

\end{document}